\documentclass[journal]{IEEEtran}
\usepackage{cite}

\usepackage{graphicx}
\usepackage{epstopdf}
\usepackage[subrefformat=parens,labelformat=parens]{subfig}

\DeclareGraphicsExtensions{.eps}
\ifCLASSINFOpdf
\else
\fi
\usepackage{amsmath}
\usepackage{amssymb}
\usepackage{amsthm}
\usepackage{algorithm}
\usepackage{algcompatible}
\usepackage{relsize}

\usepackage{tikz}
\usetikzlibrary{shapes, patterns}

\newtheorem*{theorem}{Theorem}
\newtheorem{lemma}{Lemma}

\newtheorem*{remark}{Remark}

\newcommand{\A}{\mathcal{A}}
\newcommand{\N}{\mathcal{N}}

\newcommand{\X}{\mathbf{X}}
\newcommand{\x}{\mathbf{x}}

\newcommand{\Vt}{\widetilde{V}}
\newcommand{\Pt}{\widetilde{P}}
\newcommand{\zt}{\widetilde{z}}
\newcommand{\lambdaB}{\overline{\lambda}}
\newcommand{\sigmat}{\widetilde{\sigma}}
\newcommand{\SumM}{\sum\limits^M_{m=1}}

\newcommand{\C}{\ensuremath{\mathcal{C}}}
\newcommand{\Hyp}{\ensuremath{\mathcal{H}}}

\newcommand{\LCR}{\Lambda_{\text{{\scriptsize CR}}}}

\newcommand{\LLLR}{\Lambda_{\text{{\scriptsize LLR}}}}
\newcommand{\LLFR}{\Lambda_{\text{{\scriptsize LFR}}}}

\newcommand{\LaML}{\Lambda_{\text{{\scriptsize aML}}}}
\newcommand{\E}{\ensuremath{\mathbb{E}}}
\newcommand{\Prob}{\ensuremath{\mathbb{P}}}
\newcommand{\thetab}{\boldsymbol{\theta}}

\newcommand{\lambdaOm}{\lambda_{1,m}}
\newcommand{\lambdaZm}{\lambda_{0,m}}

\newcommand{\Pois}{\text{Pois}}

\newcommand{\SNR}{\text{SNR}}
\newcommand{\lambdam}{\lambda_{1,m}}
\newcommand{\WSNfig}{
\newdimen\R
\R=1cm
\newdimen\S
\S=6cm
  
\def\h{1\S}
\def\H{2\S}
\begin{tikzpicture}[scale=0.5]

\tikzstyle {SN_style} = [ball color=white]

\tikzstyle {dSN_style} = [ball color=gray]

\tikzstyle {CH_style} = [fill = gray, very thick, shading=radial]

\draw (0,0) -- (1.8\S,0) -- (2.3\S,.9\S) -- (.5\S,.9\S) -- cycle;

\draw[dashed] (0.25\S,0.45\S) -- (2.05\S,0.45\S);
\draw[dashed] (0.9\S,0.0\S) -- (1.35\S,0.9\S);

\foreach \xcen/\ycen in { 1.8/0.5, 2.5/2, 3/1, 4.1/1.8, 4.7/0.6}
		\shade[SN_style] (\xcen, \ycen) circle (0.2cm);

\foreach \xcen/\ycen in {1.2/1.4, 4.8/2.4, 5.6/1.3}
		\shade[dSN_style] (\xcen, \ycen) circle (0.2cm);

\foreach \xcen/\ycen in {2.3/3, 3/4, 4.1/3.1, 5.4/4.3, 5.1/3.3, 6.1/3.8, 3.7/4.8}
		\shade[SN_style] (\xcen, \ycen) circle (0.2cm);

\draw (3.2,4) node[right]{$\text{SN}$};

\foreach \xcen/\ycen in {6.1/3.8, 6.5/3.0}
		\shade[dSN_style] (\xcen, \ycen) circle (0.2cm);		

\foreach \xcen/\ycen in {8.3/4.8, 9.7/4.6, 10.1/3.1, 10.7/5.1, 11.6/4.1, 12.2/5.1}
		\shade[SN_style] (\xcen, \ycen) circle (0.2cm);
		
\foreach \xcen/\ycen in {7.7/3.9, 8.5/3.3, 11.2/3.0}
	\shade[very thick, ball color=gray] (\xcen, \ycen) circle (0.2cm);

\draw[->, very thick] (7.7, 3.9) -- (0.72\S+0.9\h, 0.16\S+.45\h+\h);
\draw[->, very thick] (8.5, 3.3) -- (0.75\S+0.9\h, 0.16\S+.45\h+\h);
\draw[->, very thick] (11.2,3.0) -- (0.77\S+0.9\h, 0.16\S+.45\h+\h);		

\foreach \xcen/\ycen in {6.8/0.5, 8.2/1.3, 8.1/2.1, 9.2/2.0, 9.4/0.8, 10.1/1.6, 11.2/2.2}
		\shade[SN_style] (\xcen, \ycen) circle (0.2cm);
		
\foreach \xcen/\ycen in {7.2/1.5}
		\shade[dSN_style] (\xcen, \ycen) circle (0.2cm);
				
\tikzstyle{target}=[star, star points=5, star point ratio=2.25, draw,inner sep=0.15em,anchor=outer point 3, fill = red]
\draw[very thick] (5.7,2) node[target]{} ;
		
\draw (0,0+\h) -- (1.8\S,0+\h) -- (2.3\S,.9\S+\h) -- (.5\S,.9\S+\h) -- cycle;

\draw[dashed] (0.25\S,0.45\S+\h) -- (2.05\S,0.45\S+\h);
\draw[dashed] (0.9\S,0.0\S+\h) -- (1.35\S,0.9\S+\h);

\shade[very thick, ball color=gray] (0.55\S, 0.225\S+\h)  circle (0.4cm);
\draw[->, very thick] (0.55\S, 0.225\S+\h) -- (1.15\S, 0.40\S+\H);

\shade[very thick, ball color=gray] (0.55\S+0.9\h, 0.225\S+\h)  circle (0.4cm);
\draw[->, very thick] (0.55\S+0.9\h, 0.225\S+\h) -- (1.2\S, 0.41\S+\H);

\shade[very thick, ball color=gray] (0.75\S+0.9\h, 0.225\S+.45\h+\h)  circle (0.4cm);
\draw[->, very thick] (0.75\S+0.9\h, 0.225\S+.45\h+\h) -- (1.22\S, 0.43\S+\H);

\shade[very thick, ball color=gray] (0.75\S, 0.225\S+.45\h+\h)  circle (0.4cm);
\draw[->, very thick] (0.75\S, 0.225\S+.45\h+\h) -- (1.1\S, 0.41\S+\H);
\draw (0.68\S, 0.225\S+.45\h+\h) node[left]{$\text{CH}$};

\draw (0,0+\H) -- (1.8\S,0+\H) -- (2.3\S,.9\S+\H) -- (.5\S,.9\S+\H) -- cycle;

\shade[very thick, ball color=gray] (1.15\S, 0.5\S+\H)  circle (0.6cm);
\draw (1.15\S, 0.6\S+\H) node[above]{$\text{FC}$};

\end{tikzpicture}
}

\begin{document}
%

%
\title{Fusion Rules for Distributed Detection in Clustered Wireless Sensor Networks with Imperfect Channels}
%
%
%
\author{Sami A. Aldalahmeh,~\IEEEmembership{Member,~IEEE,}
        Saleh O. Al-Jazzar, Des McLernon, ~\IEEEmembership{Member,~IEEE,}
        Syed~Ali~Raza~Zaidi,~\IEEEmembership{Member,~IEEE,} and~Mounir Ghogho, ~\IEEEmembership{Fellow,~IEEE} \thanks{This work was supported in part by Al-Zaytoonah University of Jordan through the grant 13/28/2017-2018 and by the U.K. British Council (Newton Fund) through the grant IL3264631003 “Wireless Sensor Networks for Real-Time Monitoring of Water Quality”.} \thanks{Sami A. Aldalahmeh and Saleh O. Al-Jazzar are with the Faculty of Engineering and Technology in the Al-Zaytoonah University of Jordan, Amman Jordan (email: s.aldalahmeh@zuj.edu.jo, saleh.g@zuj.edu.jo).} \thanks{Des McLernon, Syed~Ali~Raza~Zaidi and Mounir Ghogho are with the School of Electronic and Electrical Engineering in Leeds University, Leeds, UK (email: d.c.mclernon@leeds.ac.uk, s.a.zaidi@leeds.ac.uk, m.ghogho@leeds.ac.uk).}\thanks{Mounir Ghogho is also with Department of Computer Science in Université Internationale de Rabat, Rabat, Morocco.}}
\maketitle

\begin{abstract}
In this paper we investigate fusion rules for distributed detection in large random clustered-wireless sensor networks (WSNs) with a three-tier hierarchy; the sensor nodes (SNs), the cluster heads (CHs) and the fusion center (FC). The CHs collect the SNs' local decisions and relay them to the FC that then fuses them to reach the ultimate decision. The SN-CH and the CH-FC channels suffer from additive white Gaussian noise (AWGN). In this context, we derive the optimal log-likelihood ratio (LLR) fusion rule, which turns out to be intractable. So, we develop a sub-optimal linear fusion rule (LFR) that weighs the cluster's data according to both its local detection performance and the quality of the communication channels. In order to implement it, we propose an approximate maximum likelihood based LFR (LFR-aML), which estimates the required parameters for the LFR. We also derive Gaussian-tail upper bounds for the detection and false alarms probabilities for the LFR. Furthermore, an optimal CH transmission power allocation strategy is developed by solving the Karush-Kuhn-Tucker (KKT) conditions for the related optimization problem. Extensive simulations show that the LFR attains a detection performance near to that of the optimal LLR and confirms the validity of the proposed upper bounds. Moreover, when compared to equal power allocation, simulations show that our proposed power allocation strategy achieves a significant power saving at the expense of a small reduction in the detection performance.
\end{abstract}

\begin{IEEEkeywords}
Wireless sensor networks (WSNs), cluster head (CH), distributed detection, decision fusion, power allocation strategy, KKT conditions.
\end{IEEEkeywords}
%
%
\IEEEpeerreviewmaketitle

\section{Introduction}
%
%
%
%
\IEEEPARstart{W}ireless sensor networks (WSNs) are becoming a mainstream technology constituting the backbone of several emerging technologies, such as the internet of things (IoT) \cite{Khalil2014} and smart cities \cite{Rashid2016} (see references therein). Indeed, the flexible nature of WSNs \cite{Chong2003} enables them to invade such a wide spectrum of applications. However, several aspects of WSNs remain fertile research grounds, especially distributed detection (DD) in WSNs \cite{Chamberland2007}. In such a scenario, battery-powered sensor nodes (SNs) monitor the region of interest (ROI), which are geographically distributed in a vast region in order to detect any intruders. The locations of the SNs are best modeled as a random point process \cite{Zhang2015}, because they might be out of communication range, out of power or might be even dropped from an airplane to form a network \cite{Song2009}. Due to constrained power and bandwidth, the collected data is often compressed into a single bit decision. Moreover, because the communication range is limited, providing adequate coverage for the large number of SNs is a challaging task. So, the WSN is divided into geographical clusters \cite{Bandyopadhyay2003} and hierarchically into three tiers; SNs, cluster heads (CHs) and the fusion center (FC). The SNs in each cluster send their data to the CH, which usually has more power and a larger communication range. The CHs in turn report the collected data to the FC, thus acting as high-power relays. Often data is relayed in an amplify-and-forward (AF) or decode-and-forward (DF) fashion over imperfect communication channels \cite{Hong2007}. 

In this paper, we investigate the decision fusion for distributed detection in a randomly deployed clustered-WSN operating with constrained power and over imperfect channels. In particular, the channels between SNs and CHs (termed SN-CH) and the channels between CHs and the FC (termed CH-FC) are assumed to suffer from additive white Gaussian noise (AWGN), in contrast to our previous work in \cite{Aldalahmeh2016}, which assumes ideal channels. To the best of the a authors' knowledge, this is the first work that studies fusion rules in the above problem setting.

In the light of the previous framework, the main contributions of this paper are:
\begin{enumerate}
\item The optimal log-likelihood ratio (LLR) rule is presented first, which is analytically difficult to implement. Subsequently, a sub-optimal linear fusion rule (LFR) is derived. Intuitively, the LFR  gives more weight to clusters with better detection and  channel qualities. 
\item We propose the approximate maximum likelihood estimator based fusion rule (LFR-aML) as a practical implementation of the LFR. The LFR-aML estimates the statistical parameters required for the detection fusion rule by solving a constrained maximum likelihood (ML) problem via the aid of Karush-Kuhn-Tucker (KKT) conditions.
\item To quantify the performance of the LFR and its derivatives, we derive Gaussian-tail upper bounds for the detection and the false alarm probabilities.
\item The optimal CH's transmission power is found in a closed-form manner while still adhering to a specific detection performance. This is achieved by also solving the KKT conditions of the related convex optimization problem.
\end{enumerate}

The rest of the paper is organized as follows. In Section \ref{sec:Related-Work} related work is reviewed. The adopted notation is explained in Section \ref{sec:Notation}. The system model is presented in Section \ref{sec:System-Model}. The proposed fusion rules are discussed in Section \ref{sec:Fusion-Rules}, in which we begin by formulating the optimal DD fusion rule for the noisy clustered WSN. Then, the LFR algorithm is derived in the light of the previous optimal rule. Finally, the LFR-aML is proposed as the practical implementation of the LFR. The power allocation for the LFR is investigated in Section \ref{sec:Power-Allocation}. Section \ref{sec:Practical-Consdr} discusses the practical implementation procedure of the LFR-aML algorithm. Section \ref{sec:Simulation-Result} presents the simulation results and their discussions. Finally, the conclusions are given in Section \ref{sec:Conclusion}.
\section{Related Work}
\label{sec:Related-Work}
Since the seminal work by Tenney and Sandell \cite{Tenney1981}, distributed detection has become, and still is, a rich research topic see for example \cite{Veeravalli2012}, the recent tutorial \cite{Javadi2016} and references therein. Fundamental results have been attained for parallel, tandem and tree sensor paradigm  \cite{Viswanathan1997}, emphasizing the optimality of fusing and quantizing the local log-likelihood ratios (LRTs) of the distributed sensors. Further extensions of the classical problem were presented in \cite{Blum1997}, such as weak signal detection and robust detection. 

For the case of single-bit quantization over perfect parallel networks, the opimal Chair-Varshney fusion rule (CVR) was derived in \cite{Chair1986}, which implicitly requires knowledge of the target parameters (location and power). A generalized likelihood ratio (GLRT) detector was proposed in \cite{Niu2006a} where the target parameters are estimated. However, the GLRT is computationally demanding and so the suboptimal counting rule (CR) was proposed in \cite{Niu2005}, which is simply the sum of positive local detections. Its performance on the other hand, was investigated in \cite{Niu2008}. The weighted decision fusion (WDF) is introduced in \cite{Javadi2015} as an improvement of the CR, where the decisions are weighted first before being fused at the FC. In a different direction, a detector based on the Rao test was suggested in \cite{Ciuonzo2013} and the generalized Rao test in \cite{Ciuonzo2017a} that both strike a trade-off between complexity and performance. In a similar effort, the generalized locally optimum detector was devised in \cite{Ciuonzo2017} and \cite{Ciuonzo2018}. However, the previous detectors in general suffered from the problem of spurious detection\footnote{Spurious detection is defined in \cite{Guerriero2010} as the event when a target is present and some sensors far from it declare the presence of a target in their vicinity.}, especially in a large WSN. Scan statistics-based detection was proposed in \cite{Guerriero2010} to overcome the previous problem but at the expense of a significant delay due to the sliding-window structure of the detector. A local vote decision fusion rule (LVDF) was proposed in \cite{Katenka2008}, in which sensors use neighbouring  decisions to correct their decisions locally and then integrate them globally.


DD over multi-hop (tree) sensor networks has also received considerable attention. Asymptotic results were presented in \cite{Tay2008} and \cite{Tay2009} for DD with Neyman-Pearson and Bayesian criteria, respectively. Mainly, it was shown that the error probability decays exponentially as the number of SNs increases. Decision fusion rules over multi-hop networks were investigated in \cite{Lin2005} with flat-fading noisy channels. The relaying SNs decode-and-forward the received data to the FC. The derived suboptimal rules in \cite{Lin2005} de-emphasizes sensors with more hops. Similarly, authors in \cite{Tian2007} investigated the same problem but with a binary-symmetric channel (BSC) model in the network, where it was shown that the optimal fusion rule is a weighted order statistic filter.

Clustered sensor networks were introduced for DD in \cite{Ferrari2011}, in which sensors report to  CHs that in turn report to a FC\footnote{In \cite{Ferrari2011} the CHs are refered to as FCs and the FC is refered to as an access point (AP).}. Majority-like fusion (MLF) rules were used on both the cluster level and the FC level. Surprisingly, results there show that the detection performance of a clustered sensor network is worse than the performance of sensors reporting directly to the FC. This is due to employing the MLF rule in the CHs level, thus introducing additional errors in the decision process. 

On the other hand, in our previous work \cite{Aldalahmeh2016} we have derived an optimal-cluster-based fusion rule (OCR) for clustered sensor networks. In this context, CHs collect the local SNs decisions and send this data to the FC over ideal channels. In this paper however, we consider a two-hop network in which the SN-CH and CH-FC channels are affected by AWGN. The CHs employ an AF scheme to send the collected data to the FC, in contrast to \cite{Tian2007} that adopts sending hard decisions over multiple-hop BSCs. Moreover, the adopted AF scheme enables us to minimize the transmission power, provided a specific detection performance is satisfied.




\section{Notation}
\label{sec:Notation}
In this paper we will generally refer to deterministic values by lowercase symbols $(x)$, bold symbols refers to vector values $(\x)$, whereas random values are referred to by uppercase symbols $(X)$. However, we denote the number of CHs as ($M$), global detection threshold as ($\Gamma$) and the transmitted power as ($P$). The operator $\succcurlyeq$ refers to element-wise greater than or equal to. The probability of an event $A$ is denoted by $\Prob(A)$. A normal distributed random variable (RV) $X$ with mean $\mu$ and variance $\sigma^2$ is denoted $X \sim \N(\mu,\sigma^2)$ and Poisson RV $Y$ with mean $\lambda$ is denoted by $Y \sim  \Pois(\lambda)$. The expectation operator with respect to (w.r.t.) RV $X$ is written as $\E_X[\cdot]$ and the moment generating function (MGF) for RV $X$ is defined as $M_X(t) = \E_X \left[ \exp(tX) \right]$. Finally, the estimate of any variable $x$ is denoted by $\widehat{x}$.

\section{System Model}
\label{sec:System-Model}
\begin{figure}[!ht]
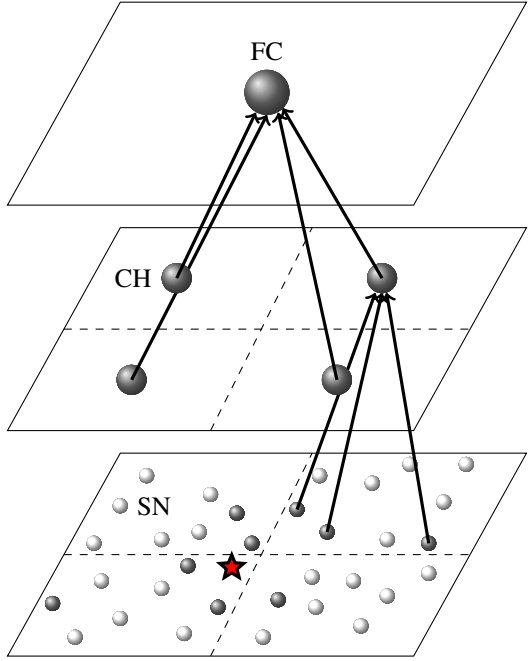

\centering
\WSNfig
\caption{The WSN topology, in which the star is the target, gray-shaded nodes are the detecting SNs and white-shaded nodes are the non-detecting SNs.}
\label{fig:WSN_fig}
\end{figure}

The WSN is functionally divided into three tiers as shown in Fig. \ref{fig:WSN_fig}. In this section we present the sensing model, the stochastic geometry model for the SNs deployment, similar to \cite{Zhang2015} and \cite{Niu2005}, and the communication model between the three tiers.
\subsection{Sensing and Sensor Deployment Models} \label{subsec:sensing-model}
Consider a WSN deployed over a region, $\A \subset \mathbb{R}^2$ where $\A$ is assumed to be significantly large. The WSN is modeled by a simple Poisson Point Process (PPP) $\Phi = \{ \X_1, \X_2, \cdots, \X_N \}$ in $\A$ \cite{Streit2010}, where $\X_i \in \Phi$ is the the coordinate of the $i$th SN. This implies that the $\X_i$'s are random variables (RVs) and their number $N = \vert \Phi \vert$ is a Poisson RV having the distribution $N \sim \Pois(\E\left[N\right])$ where $\E\left[N\right]$ is the average number of SNs. The SN intensity ($\lambda$) is defined as the average number of points (SNs) in a unit area. In general, the PPP might be non-homogeneous, i.e., the intensity is location dependent, described by $\lambda(\x)$ where $\textbf{x}$ is the location coordinates. This case might arise due to environmental or application specific constraints. However, using the non-homogeneous PPP model complicates the analysis. Thus we adopt a homogeneous PPP in our treatment ($\lambda(\x)=\lambda,\; \forall \x \in \A$), in which we can approximate the non-homogeneous case appropriately if the $m$th cluster is adequately small. In other words, the intensity in the $m$th cluster does not vary significantly with space, i.e. $\lambda_{m}(\x) \approx \lambda_{m}$ where $\lambda_m$ is the $m$th cluster SN intensity. In the homogeneous case, the number of SNs follows the distribution $\Pois(\lambda \vert \A \vert)$, where $\vert \A \vert$ is the area of $\A$. Fig. \ref{fig:WSN_fig} shows a homogeneous random network deployment.

The WSN is tasked with the detection of any intruder or target entering the ROI. A target at location $\X_t \in \A$  leaves a signature signal sensed by the SNs, which might be thermal, magnetic, electrical, seismic or  electromagnetic signal \cite{Arora2004}. We adopt the sensing model in \cite{Niu2008}, in which the signature power in the far-field  is assumed to decay quadratically with distance. The target's parameters are given in the vector $\boldsymbol{\theta} = [P_t, \X_t]^T$, where $P_t$ is the target's signal power. The noise-free signal received at the $i$th SN located at a given $\x_i$ has the following amplitude:
\begin{eqnarray}
a(\x_i) = \frac{\sqrt{P_t}}{ \max \left( d_0, d_i \right)  } \label{eq:noise-free-signal}
\end{eqnarray}
where $d_0$ is the reference distance to the node's sensor and $d_i = \| \x_t - \x_i \|$ is the Euclidean distance between the target and the $i$th SN. Note that the measured signal is saturated if the distance to the target is smaller than $d_0$. The above model can adequately describe acoustic or electromagnetic signals. 

For a given realization of $\Phi$, each SN samples the environment to decide whether an intruder is present or not. Assuming conditional independence, the collected data at the $i$th SN under the null and alternative hypotheses, $\mathcal{H}_0$ and $\mathcal{H}_1$ respectively, takes the following form:
\begin{eqnarray}
\Hyp_1: S(\x_i) &=& a(\x_i) + Q_i \\
\Hyp_0: S(\x_i) &=& Q_i 
\end{eqnarray}
where $Q_i$ is a white Gaussian noise at the $i$th SN with zero mean and variance $\sigma^2_s$. However, in practice the collected data are actually correlated \cite{Sundaresan2011}. The noise is assumed to be identically and independently distributed over all SNs and is not dependent on $\x_i$. The sensing SNR is defined as $\text{SNR}_s = P_t/\sigma^2_s$. Each SN computes its binary local decision, $I(\x_i) = \{0,1\}$, by comparing the collected data with a local decision threshold $\tau$, i.e.,
\begin{equation}
\label{eq:Ii}
I(\x_i) = \begin{cases}  1, & g\left( S(\x_i) \right) \geq \tau	\\
                     0, & g\left( S(\x_i) \right) < \tau \end{cases}
\end{equation}
where $g(\cdot)$ is the local detection function, e.g., matched filter or energy detector. Here, $\tau$ is the same for all SNs. Therefore, the local probabilities of false alarm and detection are given respectively by
\begin{eqnarray}
P_{fa} &=& f_0 \left( \tau,\sigma_s \right) \label{P_fa} \\
P_d(\x_i) &=& f_1\left( a(\x_i), \sigma_s, \tau \right) \label{eq:P_d}
\end{eqnarray}
where $f_0\left(\cdot\right)$ and $f_1\left(\cdot\right)$ are the false alarm and detection probabilities under $\Hyp_0$ and $\Hyp_1$, respectively. These functions depend on the type of local detector used, such as a matched filter or an energy detector. Note, however, that the probability of detection in \eqref{eq:P_d} also depends on the target parameters, $P_t$ and $\x_t$ through \eqref{eq:noise-free-signal}.

\subsection{Communication Model} \label{subsec:com-model}
Due to the large area of the ROI, the WSN is geographically divided into $M$ disjoint zones: $\C_1, \C_2, \cdots, \C_M $, where $\C_m \in \A$ for $m = 1, \cdots, M$. For the sake of simplicity, the $\C_m$'s are assumed to be identical. Each zone is managed by a CH located at $\x_m \notin \Phi$. The number of clusters is fixed and their locations are also fixed and known to the WSN. SNs located at $\x_i \in \C_m$ send their decisions to the $m$th CH. The CHs in turn report the collected decisions back to the FC. The three-tier network is shown in Fig. \ref{fig:WSN_fig}.

Due to cost and bandwidth constraints, SNs use on-off keying (OOK) to transmit their binary local decisions to the CH over a shared multiple access (MAC) AWGN channel. These SNs transmit with the same power $P_0$ within the cluster and are assumed to be synchronized to the same time slot. Hence, the received signal at the $m$th CH is 
\begin{equation}
	Y_m = \sqrt{P_0} \Lambda_m + W_m \label{eq:SN-CH}
\end{equation}
where  
\begin{equation}
\Lambda_m = \sum_{\X_i \in \C_m} I(\X_i), \; m = 1,\cdots,M \label{eq:Lambda_m}
\end{equation}
is the number of positive local decisions in the $m$th cluster and $W_m$ is the AWGN at that CH with distribution of $\N\left(0,\sigma^2_{c,m} \right)$. Note that since $\Lambda_m$ is actually the result of thinning of the PPP in the $m$th cluster, $\Lambda_m$ is a Poisson RV distributed as \cite{Aldalahmeh2016}
\begin{equation}
\Lambda_m \sim \begin{cases} 
					\textup{Pois} \left( \lambda_{0,m} \right),& \Hyp_0 \\
					\textup{Pois} \left( \lambda_{1,m} \right),& \Hyp_1 					
			   \end{cases} \label{eq:CR-pdf}			
\end{equation}
where $\lambda_{0,m}$ are $\lambda_{1,m}$ are the mean numbers of the detecting SNs ($\Lambda_m$) in the $m$th cluster under $\Hyp_0$ and $\Hyp_1$ respectively and are given by
\begin{eqnarray}
\lambda_{0,m} &=& \lambda P_{fa} \vert \C_m \vert \label{eq:lambda_0} \\
\lambda_{1,m} &=& \lambda \int_{\C_m} P_d(\x) d\x. \label{eq:lambda_m}
\end{eqnarray}

Note however that in the homogeneous case $\lambda_{0,m} = \lambda_0 \, \forall m$, since that $\C_m$'s are assumed to be identical.

However, in order to implement the models in \eqref{eq:SN-CH} and \eqref{eq:Lambda_m}, the CH controls the SNs' transmission power via a power control scheme. Further discussion is provided in Section \ref{sec:Practical-Consdr}. Each CH adopts the AF scheme to relay the gathered data to the FC over a dedicated AWGN wireless channel. The CHs are assumed to have more transmission power capabilities compared to the SNs. The received signals at the FC from the $m$th CH are 
\begin{equation}
Z_m = \sqrt{P_m} Y_m + V_m, \; m = 1,\cdots,M \label{eq:CH-FC}
\end{equation}
where $P_m$ is the transmission power used by the $m$th CH and $V_m \sim \N\left(0,\sigma^2_{f,m} \right)$ is the AWGN receiver in the channel between the FC and the $m$th CH.
%
\section{Fusion Rules in Clustered WSNs}
\label{sec:Fusion-Rules}
In this section we present the fusion rules for clustered WSNs. The ideal channel case is presented first as a benchmark for comparison. Then we proceed to discuss fusion rules for noisy channels. 
\subsection{Decision Fusion in Ideal Channel Clustered WSN}
For a clustered WSN with ideal communication channels, the majority-like fusion rule has been proposed in\cite{Ferrari2011}, in which the counting rule is implemented on the CH and the FC levels. However, since this rule showed a degraded performance in random WSNs so we will not discuss it further in this paper. 

In \cite{Aldalahmeh2016}, we proposed the OCR, in which CHs send the sum of the  collected SNs' decisions, $\Lambda_m$, to the FC to be optimally fused, i.e.,
\begin{equation}
\Lambda_{\text{OCR}} = \SumM c_m \Lambda_m
\label{eq:OCR} 
\end{equation}
where the optimal weighing coefficient is $c_m = \log \left( \lambda_{1,m}/ \lambda_{0,m}  \right)$. This weighing effectively suppresses the previous spurious detection problem, since the clusters containing these spurious decisions have small weighing coefficients. Note however, that $\lambda_{1,m}$ depends on the target's parameters, $\thetab$, through \eqref{eq:lambda_m}. Thus an exact implementation of \eqref{eq:OCR} requires knowledge of $\thetab$. This problem has been circumvented in \cite{Aldalahmeh2016} by using a complexity-reduced GLRT, in which $\thetab$ is coarsely estimated.

It is interesting to note however, that the CR \cite{Niu2005} is a special case of the OCR when there is only one global cluster encompassing the whole ROI. Thus the fusion rule in \eqref{eq:OCR} reduces to 
\begin{equation}
\LCR = \sum_{\X_i \in \Phi}^N I(\X_i). \label{eq:CR}
\end{equation}

The CR is also used as benchmark performance comparison of the fusion rules.
\subsection{Optimal Decision Fusion in a Noisy Clustered WSN}
In order to develop the optimal fusion rule in clustered WSNs with noisy channels, we investigate the received signals at the FC. By combining \eqref{eq:SN-CH} and \eqref{eq:CH-FC}, the received signal is
\begin{equation}
Z_m = \sqrt{\Pt_m} \Lambda_m + \Vt_m \label{eq:Z_m}
\end{equation}
where $\Pt_m = P_m P_0$ and $ \Vt_m = \sqrt{P_m} W_m + V_m$ is the aggregate noise at the $m$th CH-FC channel having a distribution of $\N\left( 0, \sigmat^2_m \right)$ where $\sigmat^2_m = P_m \sigma^2_{c,m} + \sigma^2_{f,m}$. The likelihood-ratio-test (LRT) for the signal in \eqref{eq:Z_m} is 
{\small
\begin{eqnarray}
\Lambda_{\text{LRT}} &=& \prod_{m=1}^M \frac{p\left( z_m ; \Hyp_1 \right)}{p\left( z_m ; \Hyp_0 \right)} = \prod_{m=1}^M \frac{\E_{\Lambda_m} \left[ p\left( z_m \vert \Lambda_m \right) ; \Hyp_1 \right] }{ \E_{\Lambda_m} \left[ p\left( z_m \vert \Lambda_m \right) ; \Hyp_0 \right] } \nonumber \\
  &=& \prod_{m=1}^M \frac{\E_{\Lambda_m} \left[ \exp\left( -\frac{1}{2\sigmat^2_m} \left( z_m - \sqrt{\Pt_m}\Lambda_m \right)^2 \right) ; \Hyp_1 \right] }{\E_{\Lambda_m} \left[ \exp\left( -\frac{1}{2\sigmat^2_m} \left( z_m - \sqrt{\Pt_m}\Lambda_m \right)^2 \right) ; \Hyp_0 \right]}. \nonumber \\ \label{eq:LRT}
\end{eqnarray}
}
Note that the expectations in the numerator and denominator are w.r.t. the distributions in \eqref{eq:CR-pdf}. Therefore,  $p\left( z_m ; \Hyp_1 \right)$ is actually the convolution of the Poisson distribution of $\Lambda_m$ and the Gaussian distribution of the noise leading to the fourth term in \eqref{eq:LRT}. Unfortunately, the corresponding log-likelihood ratio (LLR) is still not simpler:
\begin{eqnarray}
\LLLR &=& \sum_{m=1}^M \log \left( \E_{\Lambda_m} \left[ \exp\left( -\frac{s_m}{2} \left( \zt_m - \Lambda_m \right)^2 \right) ; \Hyp_1 \right] \right) \nonumber \\ 
&-& \log \left( \E_{\Lambda_m} \left[ \exp\left( -\frac{s_m}{2} \left( \zt_m - \Lambda_m \right)^2 \right) ; \Hyp_0 \right] \right) \label{eq:LLR-Noisy}
\end{eqnarray}
where $\zt_m = z_m / \sqrt{\Pt_m}$ and $s_m = \Pt_m / \sigmat^2_m$ is the  $m$th CH-FC channel SNR.

%
\subsection{Linear Decision Fusion Rule (LFR)}
Although the fusion rule in \eqref{eq:LLR-Noisy} is optimal, unfortunately it is impractical and does not lend itself to analysis. In order to derive a practical rule, we first need to find the distribution of $Z_m$. The MGF of $Z_m$ in \eqref{eq:Z_m} is computed as:
\begin{eqnarray}
M_{Z_m}(t) &=& M_{\Lambda_m}\left( \sqrt{\Pt_m} t \right) M_{V_m}(t) \nonumber \\
           &=& \exp\left( \lambda_{j,m} \left( e^{ t\sqrt{\Pt_m}  } -1 \right) + \frac{\sigmat^2_m}{2} t^2 \right)
\end{eqnarray}
where $j=0,1$ for $\Hyp_0$ and $\Hyp_1$ respectively. Unfortunately, the above MGF is also intractable. However, using a first order approximation of the exponential function $(e^x \approx 1+x)$ yields the following:
\begin{equation}
M_{Z_m}(t) \approx \exp \left( \lambda_{j,m} t \sqrt{\Pt_m} + \frac{\sigmat^2_m}{2} t^2 \right)
\end{equation}
which is the MGF of the Gaussian RV with $p(z_m) \sim \N \left( \lambda_{j,m} \sqrt{\Pt_m},\, \sigmat^2_m \right) $. Therefore, we approximate the LLR in \eqref{eq:LLR-Noisy} as
\begin{eqnarray}
\LLLR &\approx& \sum_{m=1}^M \frac{1}{2\sigmat^2_m} \left( z_m - \lambda_{1,m}\sqrt{\Pt_m} \right)^2  \nonumber \\
&-& \frac{1}{2\sigmat^2_m} \left( z_m - \lambda_{0,m}\sqrt{\Pt_m} \right)^2.
\label{eq:App-LLR}
\end{eqnarray}

Expanding the above and rearranging the terms gives 
\begin{eqnarray}
\LLLR &\approx& \SumM \frac{1}{2\sigmat^2_m} \left( z^2_m - 2\lambdaOm\sqrt{\Pt_m} z_m + \Pt_m\lambdaOm^2 \right) \nonumber \\
      &-& \SumM \frac{1}{2\sigmat^2_m} \left( z^2_m - 2\lambdaZm\sqrt{\Pt_m} z_m + \Pt_m\lambdaZm^2 \right) \nonumber \\
      &=& - \SumM \sqrt{\Pt_m}\left( \frac{\lambdaOm - \lambdaZm}{\sigmat^2_m} \right) z_m  \nonumber \\
      &+& \SumM \Pt_m\left( \lambdaOm^2 + \lambdaZm^2 \right). 
\label{eq:LFR-long}
\end{eqnarray}


When comparing $\LLLR$ with the detection threshold ($\Gamma$), the last term in \eqref{eq:LFR-long} is absorbed by $\Gamma$ since it is independent of $z_m$ .  So the resulting linear fusion rule (LFR) becomes:
\begin{equation}
\LLFR = \sum_{m=1}^M d_m z_m \gtrless \Gamma, \label{eq:LFR}
\end{equation}
\label{eq:Noisy-FR}
where the linear weighing coefficients are
\begin{equation}
d_m = \frac{\sqrt{\Pt_m}}{\sigmat^2_m}\left( \lambda_{1,m} - \lambda_{0,m} \right). \label{eq:dm}
\end{equation}

The LFR is essentially a weighted sum of the data provided by each cluster. The impact of each cluster is reflected by its weight $d_m$, which is a measure of the detection performance and the channel quality of that cluster. 

\begin{remark}
The LFR intuitively gives more weight to clusters with better detection, which is manifested in the mean difference term $\left( \lambda_{1,m} - \lambda_{0,m} \right)$. Also, more weight is given to clusters with good channel quality, i.e., large $\sqrt{\Pt_m}/\sigmat^2_m$.
\end{remark}

Clearly the LFR is computationally simple, in contrast to the LLR in \eqref{eq:LLR-Noisy}. In fact, its computational complexity amounts to $O(M)$ only.
\subsection{Approximated Maximum Likelihood-Based LFR (LFR-aML)}
Although the LFR is a analytically simple, its implementation requires $\lambda_{1,m}$'s to be known by the FC. Hence, we extend the LFR in this section to include an estimation phase prior to the detection phase.

\subsubsection{Estimation phase}
At first glance, the $\lambdam$'s can be computed by initially estimating $\thetab$ through maximum likelihood estimation from the log-likelihood of $Z_m$'s under $\Hyp_1$ (the first term in \eqref{eq:LLR-Noisy}). However, such an estimation problem is also nonlinear and nonconvex, leading to high computational complexity. So we propose estimating $\lambda_{1,m}$'s directly. Still, attempting to do that from the current log-likelihood expression will not provide satisfactory results since each CH provides a single data point about $Z_m$, consequently, several instances of $Z_m$ are required. Thus we extend the LFR to the multiple-sample case, in which each SN makes $L$ independent decisions, $\lbrace I_{i,l} \rbrace_{l=0}^{L-1}$, that are relayed to the FC by the CHs. Further details of the implementation are provided in Section \ref{sec:Practical-Consdr}.

Given the set of collected data $\widetilde{z}_{l,m}$'s, which is the $l$th sample from the $m$th CH, then the corresponding likelihood function is $\prod_{m=1}^M \prod_{l=0}^{L-1} p\left( \widetilde{z}_{l,m} ; \lambda_{1,m} \right)$. It follows directly that the constrained ML problem using the related log-likelihood is 
\begin{eqnarray}
\label{eq:ML}
&& \max_{\lambdam} \, \sum_{m=1}^M \sum_{l=0}^{L-1} \log \left( p\left( \widetilde{z}_{l,m} ; \lambda_{1,m} \right) \right)\\ 
&& \text{s.t.} \quad \lambdam \geq \lambda_0 \quad \forall m.  \nonumber
\end{eqnarray}
%

Even though the ML problem above is separable in $\lambda_{1,m}$, it is still complicated. Therefore, we propose to solve a suboptimal version of \eqref{eq:ML} by using the corresponding lower bound provided by the following lemma:

\begin{lemma}
\label{lem:Bound}
The lower bound of the log-likelihood function in \eqref{eq:ML} is given as
\begin{equation}
\label{eq:ML_bound}
\log\left( p\left( \widetilde{z}_{l,m} ; \lambda_{1,m} \right) \right) \geq  \widehat{\Lambda}_{l,m}\log\lambda_{1,m} -\lambda_{1,m} + C_1
\end{equation}
where 
\begin{equation}
\widehat{\Lambda}_{l,m} = \frac{\sum\limits_{k=0}^{\infty} k p\left( \widetilde{z}_{l,m} | k \right)}{\sum\limits_{k=0}^{\infty} p\left( \widetilde{z}_{l,m} | k \right)}
\label{eq:zm_hat}
\end{equation}
is the mean estimate of the $l$th received sample from the $m$th CH and $p\left( \widetilde{z}_{l,m} | k \right)$ is the conditional Gaussian distribution given $k$ and 
\begin{equation}
C_1 = \log C_0 - \sum_{k=0}^{\infty} \pi_k \log k!
\end{equation}
where $C_0 = \sum_{k=0}^{\infty}  p\left( \widetilde{z}_{l,m} | k \right)$ and $\pi_k = p\left( \widetilde{z}_{l,m} | k \right) /C_0$.
\end{lemma}
\begin{proof}
See Appendix \ref{app:App-Lemma}.
\end{proof}

Consequently, we have the surrogate optimization problem:
\begin{eqnarray}
\label{eq:app-ML}
&&\max_{\lambdam} \, \sum_{m=1}^M \sum_{l=1}^L \left( \widehat{\Lambda}_{l,m} \log\lambdam - \lambdam \right) \\
 &&\text{s.t.} \quad  \lambdam \geq \lambda_0 \quad \forall m. \nonumber
\end{eqnarray}

The optimal solution is given in the following lemma.

\begin{lemma}
\label{lem:const-opt}
The optimal solution of the constrained optimization problem in \eqref{eq:app-ML} is

\begin{eqnarray}
\widehat{\lambda}_{1,m} &=& \begin{cases} 
				 \widehat{\Lambda}_{m}, \quad \eta_m < 0  \\
				 \lambda_0, \quad \eta_m = 0
			\end{cases} \label{eq:lambda_m_est} \\	
\eta_m &=& 1 - \widehat{\Lambda}_{m}/\lambda_0
\label{eq:eta}	
\end{eqnarray}
where 
\begin{equation}
\label{eq:Lambda-ave}
\widehat{\Lambda}_{m} = \frac{1}{L}\sum_{l=0}^{L-1} \widehat{\Lambda}_{l,m}
\end{equation}
is the average of the $m$th CH's estimates $\widehat{\Lambda}_{l,m}$ given in \eqref{eq:zm_hat}.

\end{lemma}

\begin{proof}
See Appendix \ref{app:App-Lemma-const-opt}.
\end{proof}

The estimator $\widehat{\lambda}_{1,m}$ in \eqref{eq:lambda_m_est} is intuitive in the sense that it is the average of all sample estimates when a target is sensed and is simply $\widehat{\lambda}_{1,m} = \lambda_0$ otherwise.
\subsubsection{Detection phase}
Given the estimates $\widehat{\lambda}_{1,m}$'s, we then propose the LFR-aML detector as
\begin{equation}
\LaML = \sum_{m=1}^M \widehat{d}_m \widehat{z}_{m} \gtrless \Gamma
\label{eq:aML}
\end{equation}
where the weighing coefficients now are
\begin{equation}
\widehat{d}_m = \frac{\sqrt{\Pt_m}}{\sigmat^2_m}\left( \widehat{\lambda}_{1,m} - \lambda_{0} \right) \label{eq:d_m_hat}
\end{equation}
and the $\widehat{z}_m$ is the averaged data for each CH defined as
\begin{equation}
\label{eq:z_m_ave}
\widehat{z}_m = \frac{1}{L} \sum_{l=1}^L \widetilde{z}_{l,m}.
\end{equation}

The computational complexity of the LFR-aML, on the other hand, is $O\left( L M \right)$, which is relatively larger than that for the LFR. 
\subsection{Linear Fusion Rule Performance}
Despite being a practical fusion rule, the LFR's  detection and false alarm probabilities do not have closed forms that lend themselves to analysis. Thus we resort to finding the upper bound for those tail probabilities in the following theorem.
\begin{theorem}
The upper bound for tail probability for the LFR is 
\begin{equation}
\Prob \left( \LLFR > z ; \Hyp_j \right) \leq \exp \left( - \frac{\left( z - \lambdaB_{j,d} \right)^2}{2\sigmat^2_{j,d}} \right)
\label{eq:Thrm-Bnd}
\end{equation}
\label{thrm:Noisy-Mod-Chrnf-Bound}
for $j=0,1$, where
\begin{eqnarray}
\lambdaB_{j,d} &=& \SumM \lambda_{j,m} d_m \sqrt{\Pt_m} \label{eq:lambda_bar} \\
\sigmat^2_{j,d} &=& \SumM d^2_m \left( \lambda_{j,m} \Pt_m + \sigmat^2_m \right)\label{eq:sigma2_bar}.
\end{eqnarray}
\end{theorem}
\begin{proof}
See Appendix \ref{app:App-Noisy-Mod-Chrnf-Bound}.
\end{proof}

Clearly, the upper bound given above is a Gaussian-tail bound. It is interesting to note that the mean defined in \eqref{eq:lambda_bar} is actually a scaled version of the $\LLFR$ mean defined in \eqref{eq:LFR}. Thus it is expected that if the LFR value is increased the detection probability will significantly improve.
\section{Power Allocation}
\label{sec:Power-Allocation}
WSNs are notorious for being power constrained. Hence, the power should be used wisely, especially if the application is critical such as intruder detection,. It is well established that the main source of power usage in a WSN is wireless communication \cite{Raghunathan2002}. Thus it is desired to minimize the transmission power used by SNs and CHs while jointly taking into consideration the minimum required detection performance. 

Fortunately, the linear LFR structure facilitates the use of power allocation strategy, in contrast to the LLR. The mean-difference (MD) \cite{Liu2005b} is adopted  as the detection performance criteria due to its desirable form. The MD is defined as 
\begin{eqnarray}
\text{MD} &=& \lambdaB_{d,1} - \lambdaB_{d,0} \nonumber \\
&=& \SumM d_m \sqrt{\Pt_m} \left( \lambda_{1,m} - \lambda_0  \right) \nonumber \\
          &=& \SumM \frac{P_0 P_m \left( \lambda_{1,m} - \lambda_0  \right)^2}{\sigma^2_{c,m} P_m + \sigma^2_{f,m}}.
\end{eqnarray}

In general the MD and the detection performance can be significantly improved by increasing the SNs deployment density, $\lambda$, without the need to increase the transmission power. So, attention should be directed at reducing the transmission power given a specific detection performance constraint\footnote{Although $P_0$ is responsible for a large part of the used power, the $P_m$'s on the other hand, play a more critical role in the WSN since if any CH runs out of power a significant part of the network is rendered useless}. Thus we wish to solve the following optimization problem:
\begin{eqnarray}
\label{eq:Pwr-Allc}
\min_{\mathbf{P}} && \Vert \mathbf{P} \Vert_1 \\
\text{s.t.} \quad \mathbf{P} & \succcurlyeq & \mathbf{0} \nonumber \\
 \quad \text{MD} &=& P_0 \SumM \frac{ P_m \left( \lambda_{1,m} - \lambda_0  \right)^2}{\sigma^2_{c,m} P_m + \sigma^2_{f,m}} > D_0 \nonumber
\end{eqnarray}
where $\mathbf{P} = \left(P_1, P_2, \cdots, P_M \right)^T$ is the lumped CH's transmission powers vector and $D_0$ is the minimum mean difference as specified by the network. The $l_1$-norm is adopted since it reduces the residuals leading to smaller component values in $\mathbf{P}$ and hence less transmission power. 

Note however, that the above power allocation problem \eqref{eq:Pwr-Allc} is very similar to the water-filling problem \cite{Boyd2004}, but here the power sum is minimized in contrast to the conventional water-filling formulation. Obviously, the above problem is convex, since the objective function is convex and the constraint is a linear-fractional function, which is also convex \cite{Boyd2004}. However, the $l_1$-norm is not differentiable and consequently a closed form solution cannot be attained. So we replace the $\Vert \mathbf{P} \Vert_1$ by the summation of the elements of $\mathbf{P}$'s (where each element is non-negative).  

\begin{theorem}
\label{thrm:Opt-pwr-alloc}
The optimal power allocation based on the formulation in \eqref{eq:Pwr-Allc} is given as

\begin{equation}
P_m = \left( \dfrac{  \lambda_{d,m} \sigma_{f,m} \sqrt{\nu}}{ \sigma^2_{c,m}} - \dfrac{\sigma^2_{f,m}}{\sigma^2_{c,m}} \right)^{+} \label{eq:Pm}
\end{equation}
where $(x)^+ = \max(0,x)$ and
%
\begin{equation}
\sqrt{\nu} = \dfrac{\mathlarger{\SumM} \dfrac{\lambda_{d,m} \sigma_{f,m}}{\sigma^2_{c,m}}}{ \mathlarger{\SumM} \dfrac{\lambda^2_{d,m}}{\sigma^2_{c,m}} - D_1}.
\label{eq:nu}
\end{equation}

\end{theorem}

\begin{proof}
See Appendix \ref{app:Opt-pwr-alloc}.
\end{proof}
 
%

Intuitively, the allocated power is proportional to the cluster's detection performance manifested in the mean difference $\lambda_{d,m}$ and is inversely proportional to the SN-CH channel noise, $\sigma^2_{c,m}$. Of course, when using the power allocation algorithm practically, $\widehat{\lambda}_{1,m}$ is used to compute $\lambda_{d,m}$.

In this work, we will denote the LFR with power allocation strategy as LFR-PA, whereas the LFR using equal power allocation is just denoted by LFR. Similarly, the fusion rule using the estimates in \eqref{eq:eta} and \eqref{eq:lambda_m_est} is called as LFR-aML.

\section{Practical Considerations}
\label{sec:Practical-Consdr}
In this section we discuss how the LFR-aML algorithm is implemented in practice.   

The LFR-aML is preceded by an initialization stage where the communication parameters are estimated. In order to ensure a constant received power at the CHs, thus validating the model in \eqref{eq:Lambda_m}, a simple power control scheme is implemented. The CHs send pilot signals to the SNs in the cluster to be used to adapt the SNs' transmission power accordingly.  Moreover, the CHs compute the SN-CH channel noise variances, $\sigma^2_{c,m}$, and likewise the FC computes the CH-FC channel noise  variances, $\sigma^2_{f,m}$. Finally, $\lambda_0$ can be estimated off-line.

Then the LFR-aML is initiated where it performs estimation of the clusters' average number of detecting SNs, global distributed detection and optimal CH power allocation. Algorithm \ref{alg:LFR_aML} illustrates the complete LFR-aML detection procedure.

%
%

\begin{algorithm}[t]
\caption{: LFR-aML}	
\label{alg:LFR_aML}
	\textbf{Initialization:}
	\begin{algorithmic}[1]
		\STATE CHs send pilot signals to SNs.
		\STATE SNs use pilot signal to adjust $P_0$.		
		\STATE CHs estimate the $\lbrace\sigma^2_{c,m}\rbrace_{m=1}^M$.
		\STATE FC estimate the $\lbrace\sigma^2_{f,m}\rbrace_{m=1}^M$.
		\STATE FC estimates $\lambda_{0}$ via \eqref{eq:lambda_0}, where $\lambda_{0,m} = \lambda_0\, \forall m$.
	\end{algorithmic}
	\textbf{LFR-aML:}	
	\begin{algorithmic}[1]
	\STATE $P_m = P_{tot}/M$.
		\LOOP
		\STATEx \textbf{Estimation:}
		\STATE SNs compute $\lbrace I_{i,l} \rbrace_{l=0}^{L-1}$ via \eqref{eq:Ii}.
		\STATE SNs send $\lbrace I_{i,l} \rbrace_{l=0}^{L-1}$ to the CHs.
		\STATE CHs compute $\lbrace \widehat{\lambda}_{1,m} \rbrace_{m=1}^M$ via \eqref{eq:lambda_m_est}, \eqref{eq:eta} and \eqref{eq:Lambda-ave}.
		\STATE CHs compute $\lbrace\widehat{z}_m\rbrace_{m=1}^M$ via \eqref{eq:z_m_ave}.
		\STATE CHs use $P_m$ to send $\lbrace \widehat{\lambda}_{1,m},\widehat{z}_m\rbrace_{m=1}^M$ to FC.		
		\STATEx \textbf{Detection:}
		\STATE FC computes $\LaML$ via \eqref{eq:aML}, \eqref{eq:d_m_hat} and \eqref{eq:z_m_ave}.		
		\STATE FC tests condition $\left(\LaML  \gtrless \Gamma\right)$ for global detection.
		\STATEx \textbf{Power Allocation:}
		\STATE FC computes $P_m$'s via \eqref{eq:Pm} and \eqref{eq:nu}.
		\STATE FC sends $P_m$'s to CHs.	
		\ENDLOOP
	\end{algorithmic}
	
\end{algorithm}
%
%
%
%
%
\begin{figure*}[!ht]
\centering
\subfloat[$\SNR_{f,m} = \SNR_{c,m} = 5,\text{dB}, \, \forall m.$ \label{fig:ROC-SNR-Eq-hi}]{ \includegraphics[width=.45\textwidth]{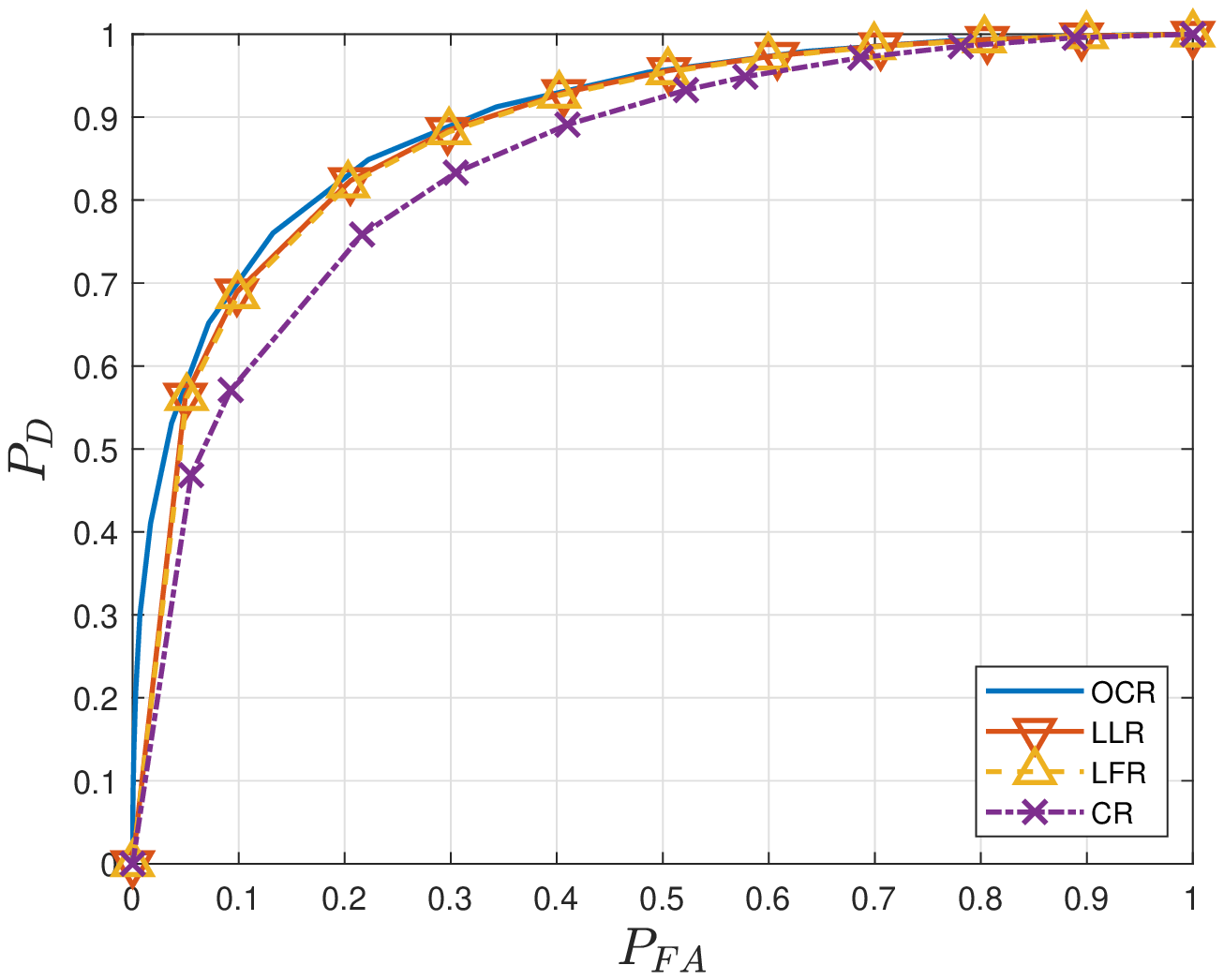}} \hfill
\subfloat[$\SNR_{f,m} = \SNR_{c,m} = -5\,\text{dB}, \, \forall m.$\label{fig:ROC-SNR-Eq-lo}]{ \includegraphics[width=.45\textwidth]{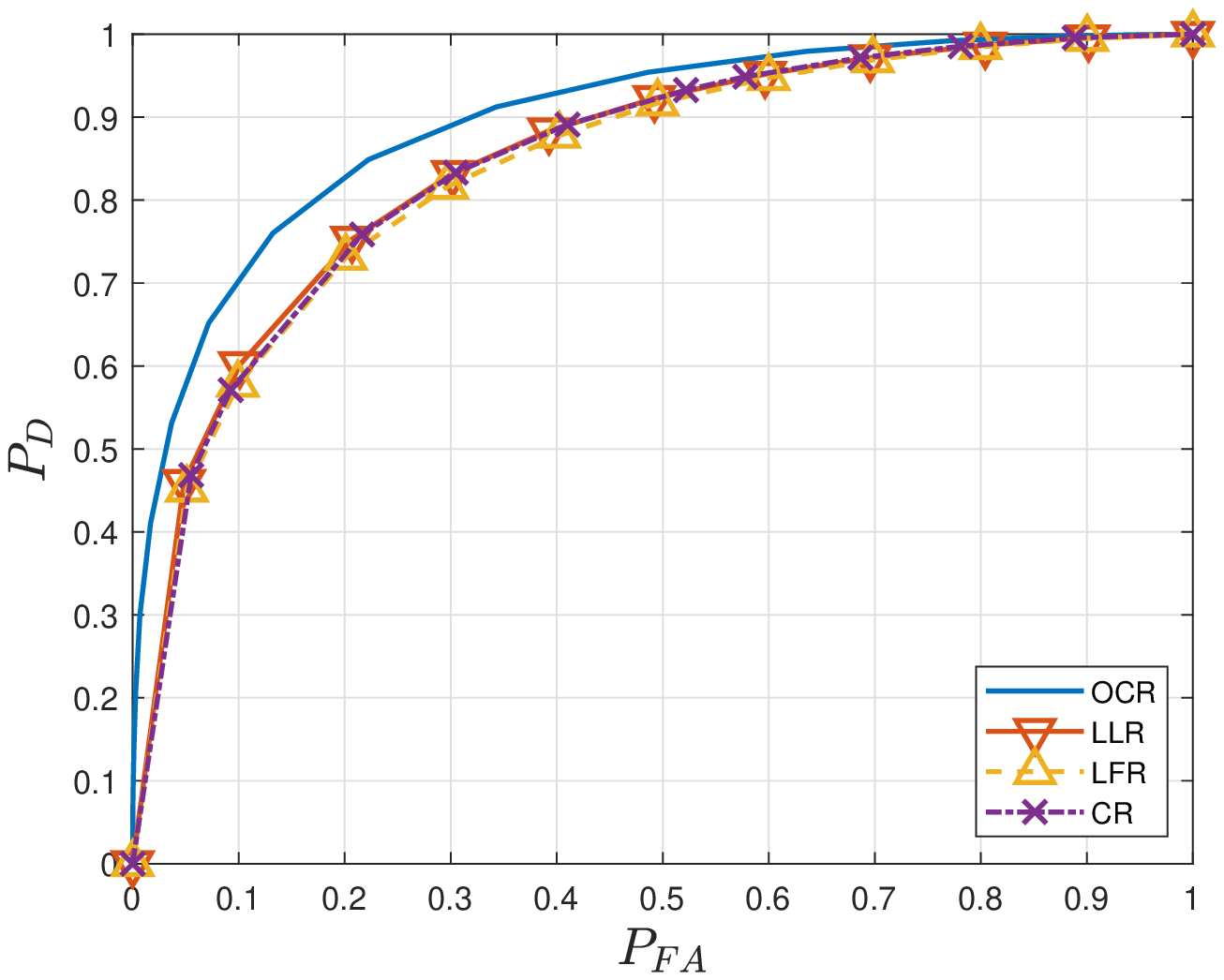}} \\
\subfloat[$\SNR_{f,m} = 0\,\text{dB} < \SNR_{c,m} = 5\,\text{dB}, \, \forall m.$\label{fig:ROC-SNR-Cm-gt-Fm}]{ \includegraphics[width=.45\textwidth]{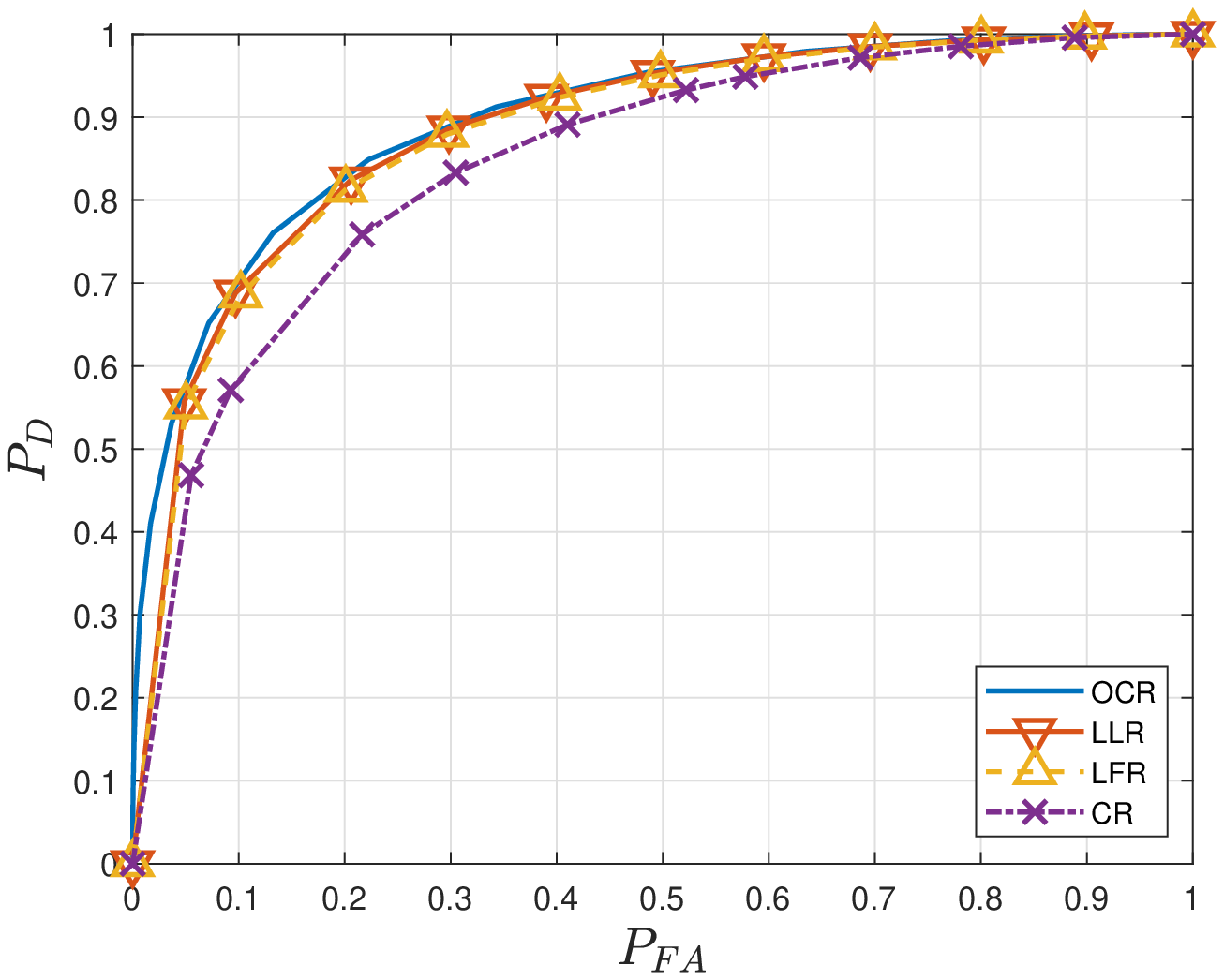}}\hfill
\subfloat[$\SNR_{f,m} = 5\,\text{dB} > \SNR_{c,m} = 0\,\text{dB}, \, \forall m.$\label{fig:ROC-SNR-Fm-gt-Cm}]{ \includegraphics[width=.45\textwidth]{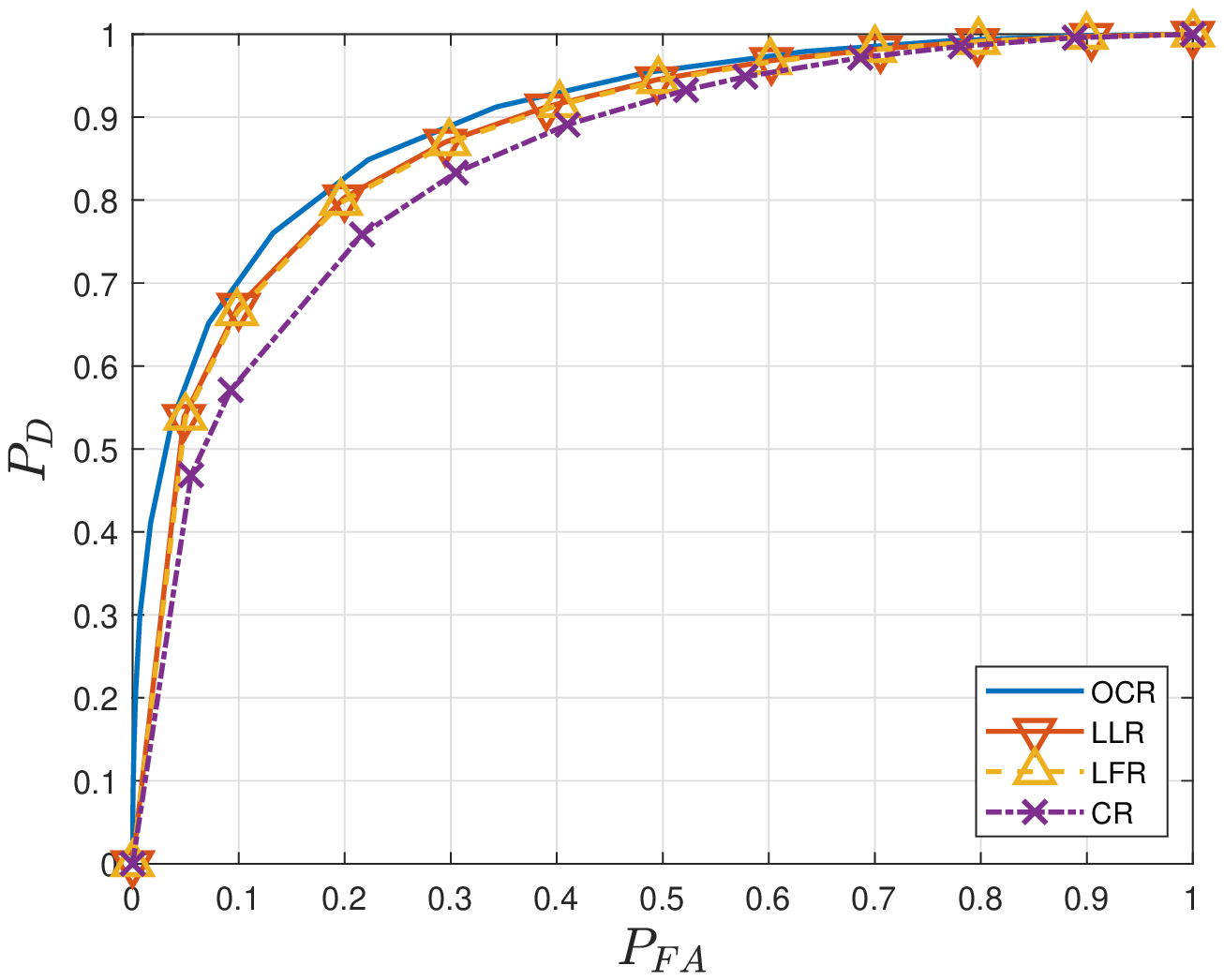}} 
\caption{Effect of SN-CH and CH-FC channel SNRs on the detection performance. The SN transmission power is, $P_0 = 1$ and the CH transmission power is $P_m = 1,\, \forall m$.}
\label{fig:ROC-SNR}
\end{figure*}
%
%
%
\begin{figure*}[t]
\centering
\subfloat[Cluster containing target has $\SNR_{c,m} = 5\,\text{dB}$, rest of cluster have $\SNR_{f,m} = \SNR_{c,m} = 2,\text{dB}.$\label{fig:ROC-SNR-Target-hi-SNR_c}]{\includegraphics[width=.45\textwidth]{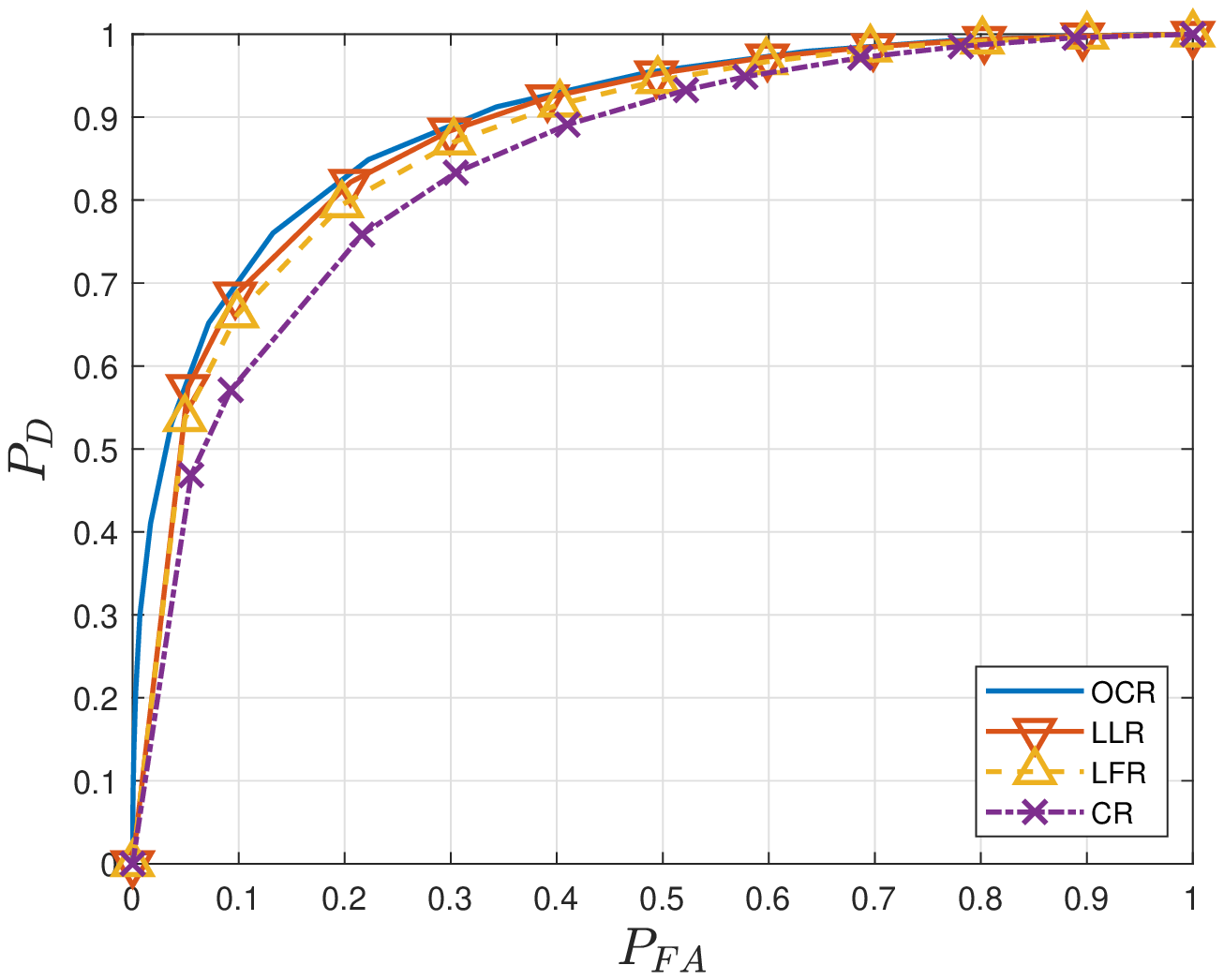}}\hfill
\subfloat[Cluster containing target has $\SNR_{c,m} = -5\,\text{dB}$, rest of cluster have $\SNR_{f,m} = \SNR_{c,m} = 2,\text{dB}.$\label{fig:ROC-SNR-Target-lo-SNR_c}]{\includegraphics[width=.45\textwidth]{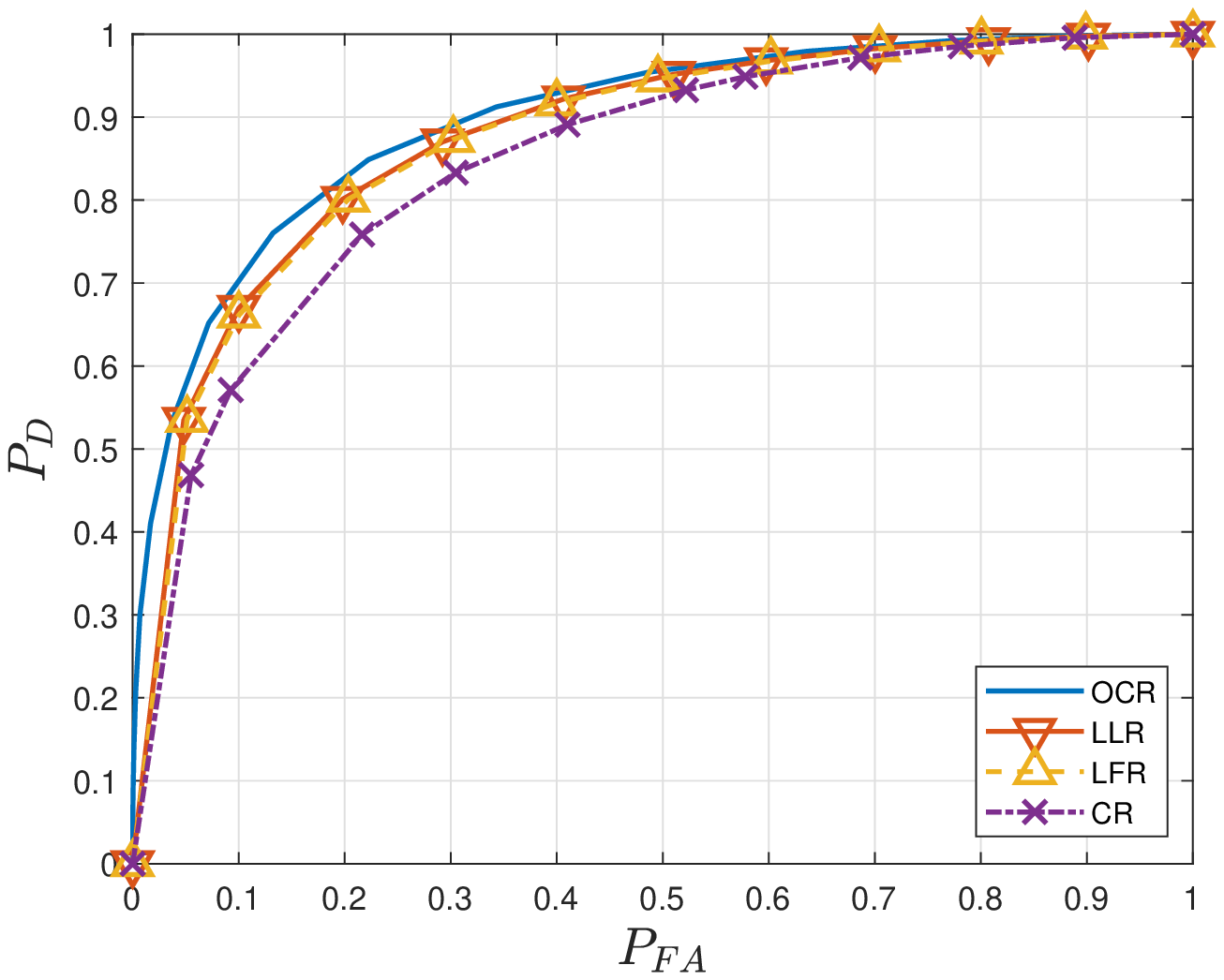}} \\
\subfloat[Cluster containing target has $\SNR_{f,m} = 5\,\text{dB}$, rest of cluster have $\SNR_{f,m} = \SNR_{c,m} = 2,\text{dB}.$\label{fig:ROC-SNR-Target-hi-SNR_f}]{\includegraphics[width=.45\textwidth]{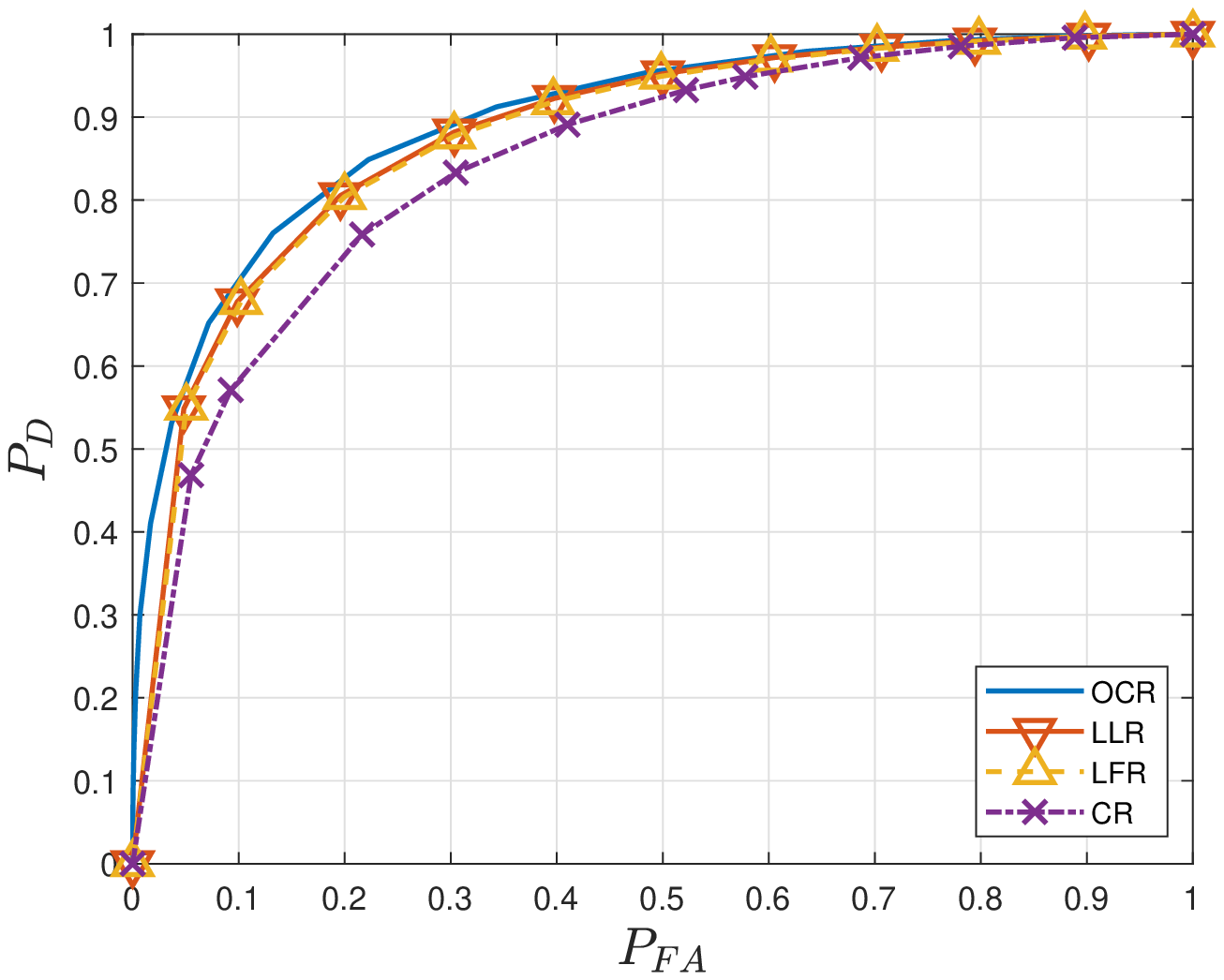}}\hfill
\subfloat[Cluster containing target has $\SNR_{f,m} = -5\,\text{dB}$, rest of cluster have $\SNR_{f,m} = \SNR_{c,m} = 2,\text{dB}.$\label{fig:ROC-SNR-Target-lo-SNR_f}]{\includegraphics[width=.45\textwidth]{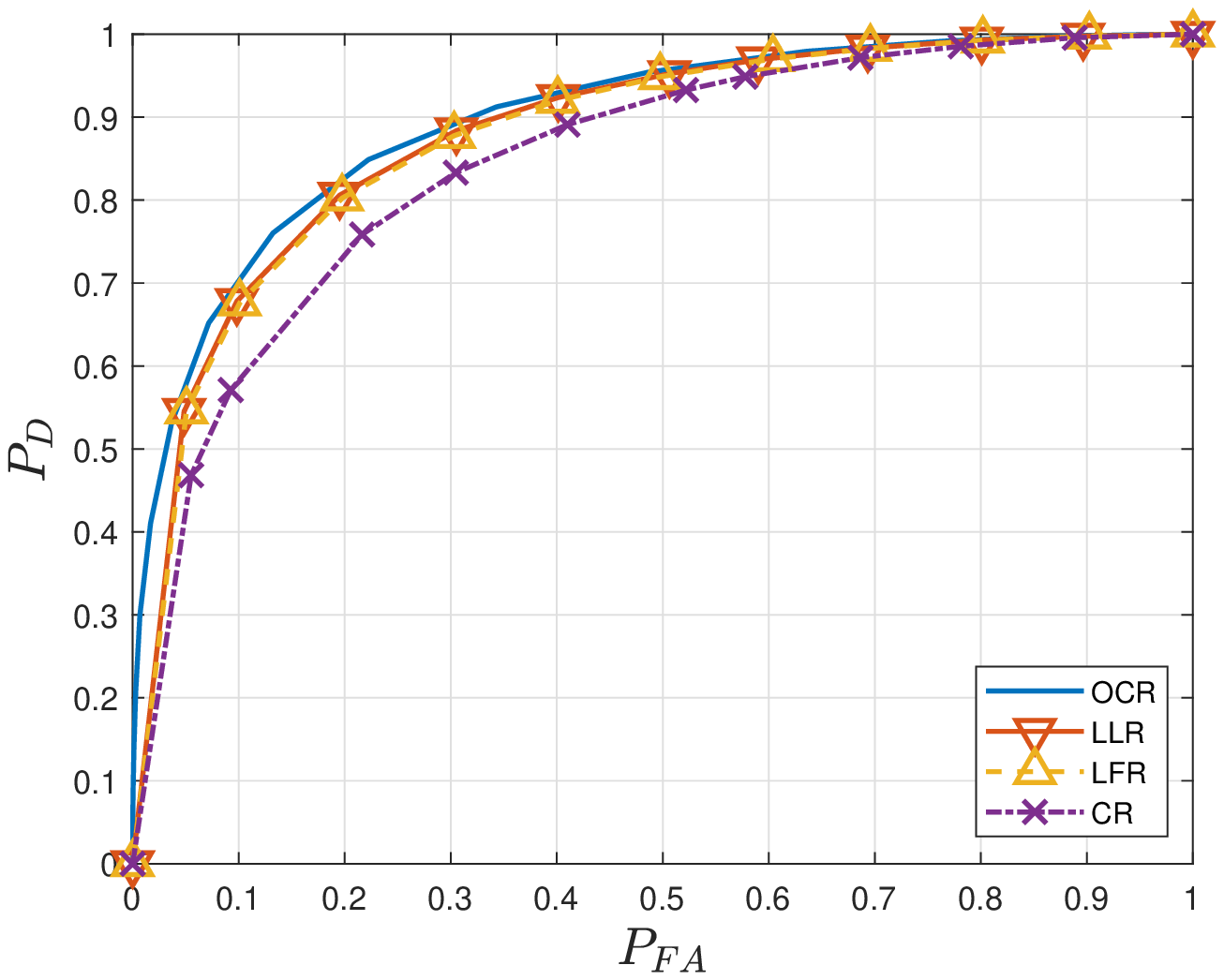}}
\caption{Effect of SN-CH and CH-FC channel SNRs on the detection performance. The SN transmission power is $P_0 = 1$ and the CH transmission power is $P_m = 1,\, \forall m$.}
\label{fig:ROC-SNR-Target}
\end{figure*}
%
%
%
\begin{figure*}[t]
\centering
 \subfloat[$P_{FA}$ Bound.\label{fig:PFA-Bnd}]{ \includegraphics[width=.45\textwidth]{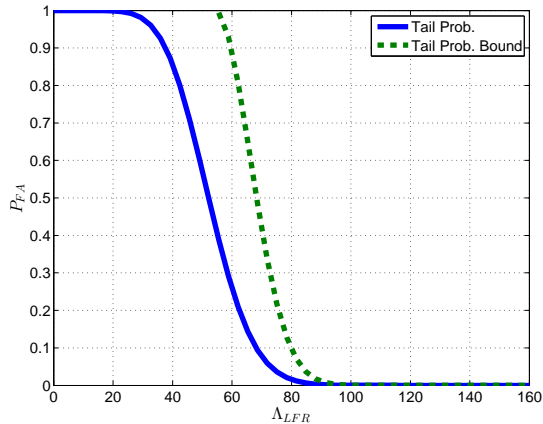}}
 \hfill
 \subfloat[$P_D$ Bound.\label{fig:PD-Bnd}]{ \includegraphics[width=.45\textwidth]{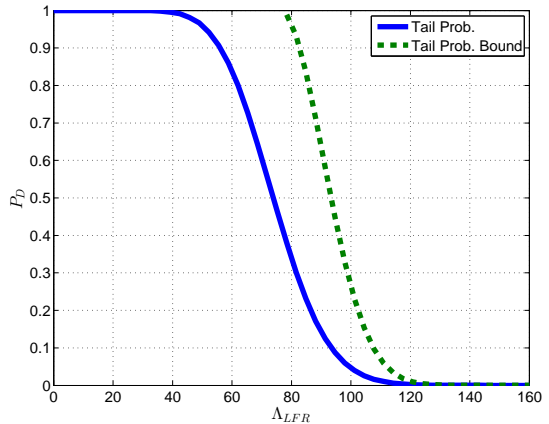}}

\caption{Tail Bounds for  false alarm probability and detection probability. The SNs transmission power is $P_0 = 1$ and the CHs transmission power is $P_m = 1,\, \forall m$.}
\label{fig:Tail-Prob-Bnd}
\end{figure*}
%
%
%
\begin{figure*}[t]
\centering
 \subfloat[ROC.\label{fig:ROC_aML}]{ \includegraphics[width=.45\textwidth]{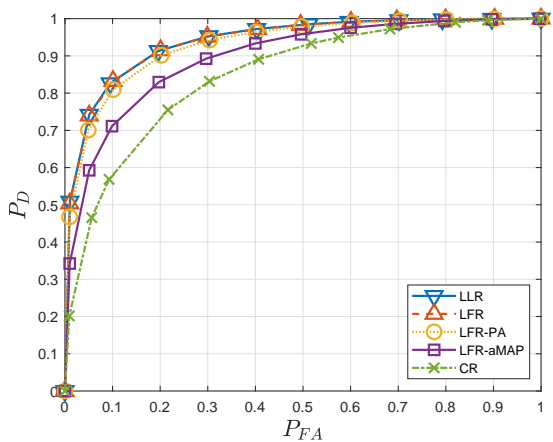}}
 \hfill
 \subfloat[Power used $P_m$ by CHs.\label{fig:PWR_aML}]{ \includegraphics[width=.45\textwidth]{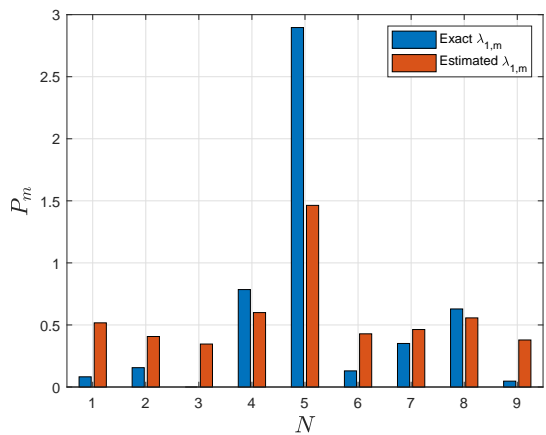}}
\caption{The effect of using the aML estimator on the ROC and the power used at $P_0 = 2$, data samples of $L=5$, $\SNR_{f,m} = 2 \text{ and } \SNR_{c,m} = 2,\text{dB}, \, \forall m.$}
\label{fig:ROC_PWR}
\end{figure*}
%
%
%
\begin{figure*}[t]
\centering
 \subfloat[$P_D$ at $P_{FA} = 0.1$.\label{fig:PD_K1_FC_CH_2}]{ \includegraphics[width=.45\textwidth]{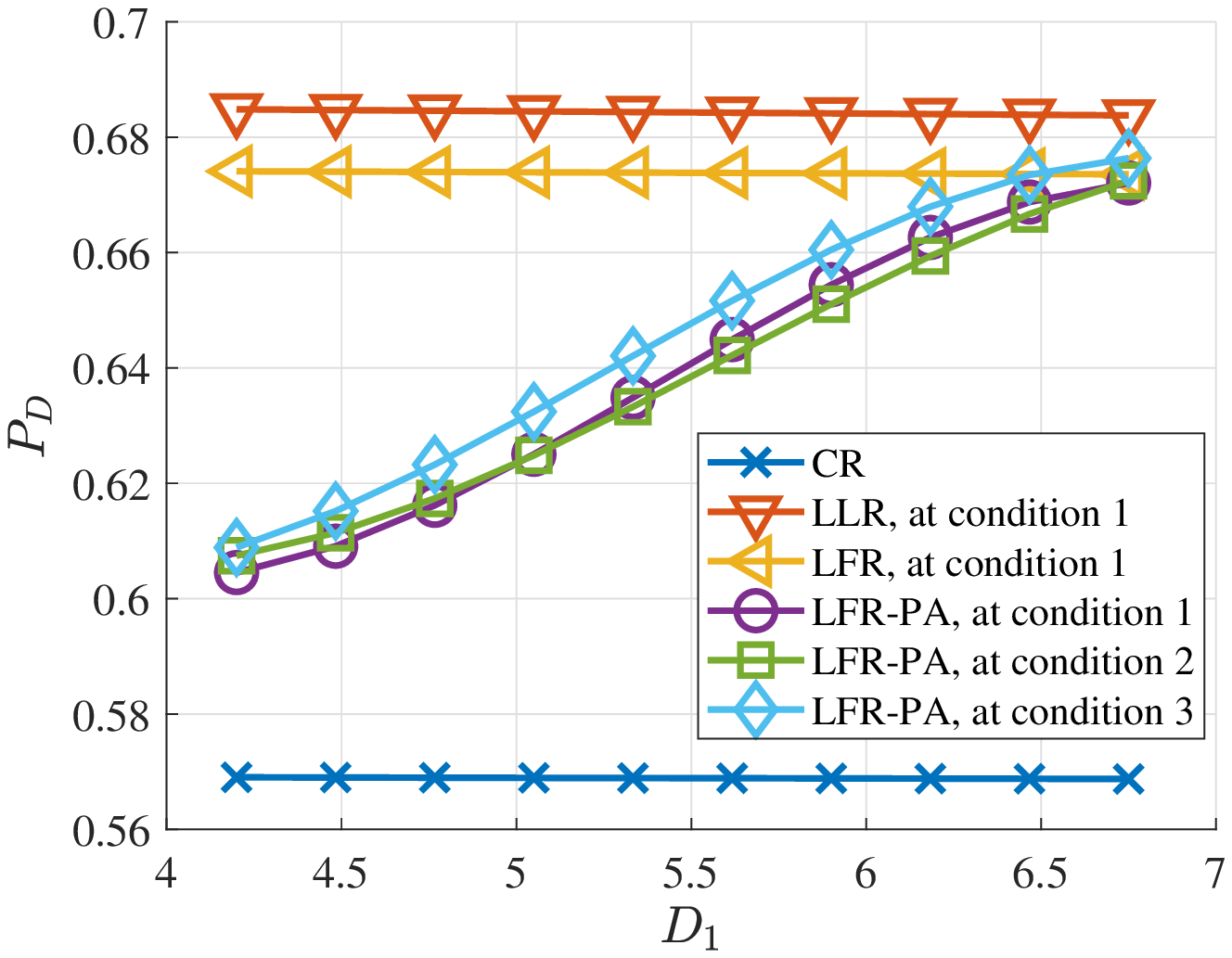}}
 \hfill
 \subfloat[Saved power percentage compared to the LFR rule.\label{fig:Psav_K1_FC_CH_2}]{ \includegraphics[width=.45\textwidth]{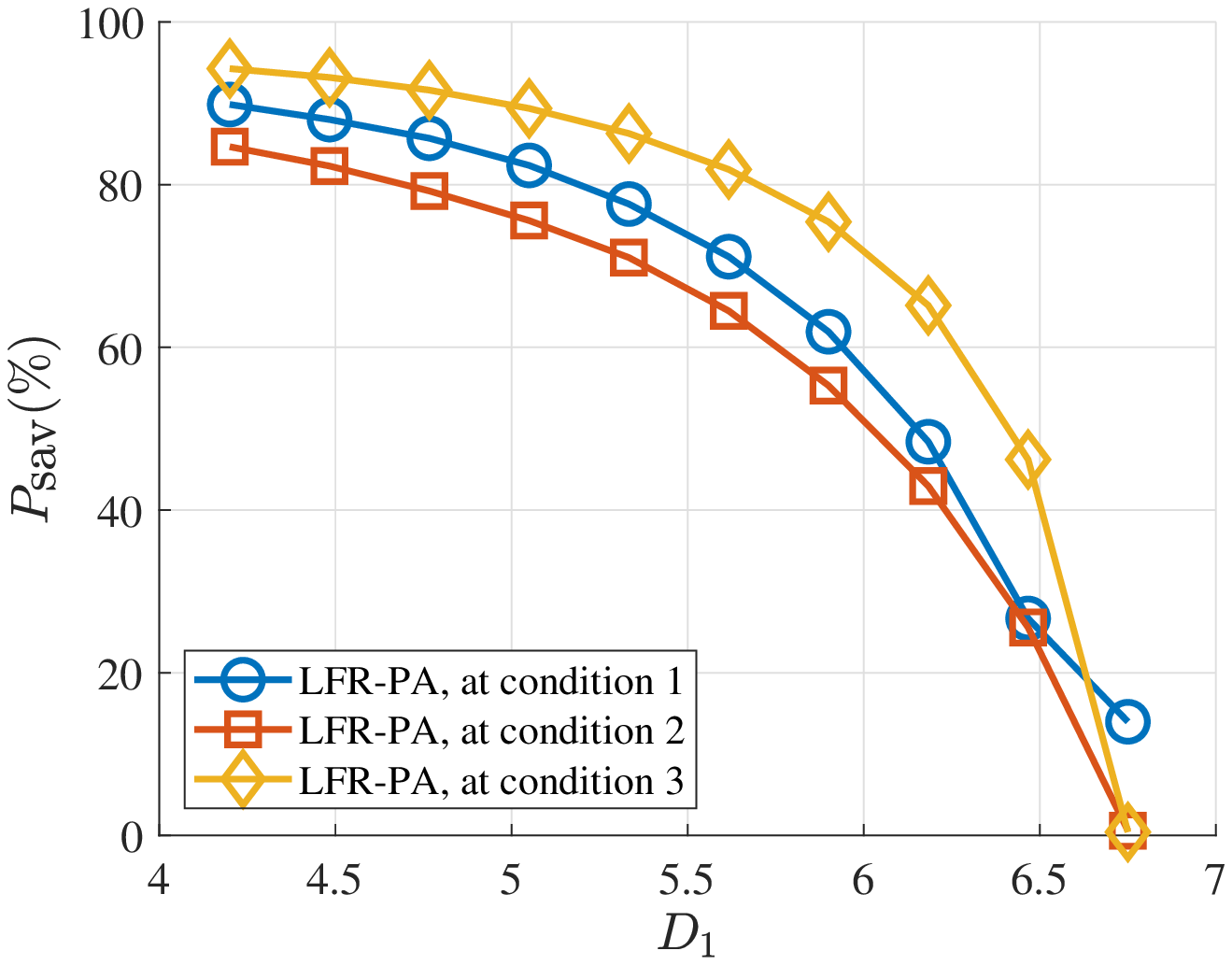}}
\caption{Detection performance and the power saving percentage achieved using the LFR-PA rule at $P_0 = 5$ and $P_m = 2$ and the following conditions: (1) $\SNR_{f,m} = \SNR_{c,m} = 2\text{dB}$, (2) $\SNR_{f,m} = 2$ and $\SNR_{c,m} = 5\text{dB}$ and (3) $\SNR_{f,m} = 5$ and $\SNR_{c,m} = 2\text{dB}$.}
\label{fig:PD_Psav_K1_FC_CH_2}
\end{figure*}
\section{Simulation Results and Discussion}
\label{sec:Simulation-Result}
We simulate a WSN deployed in a $50 \times 50$  ROI. The intruder's power is $P_0 = 1$ located arbitrarily at $(4,5)$. The sensing SNR is set to 0 dB. The SNs have a reference distance of $d_0 = 1$ units with a local probability of false alarm of $10^{-2}$. The SNs adopt the matched filter as their local detector. The WSN has a SN deployment density of $\lambda=2$ per unit area and is divided into 9 clusters. The system is simulated for $10^5$ Monte Carlo iterations. Note however, that the simulation setting here is arbitrary but these results also hold for different scenarios.

Fig.\,\ref{fig:ROC-SNR} shows the effect of the SNRs of the SN-CH and the CH-FC channels on the detection performance through the ROC graphs. The OCR and CR are included as upper and lower bounds for the proposed algorithms. The case of having equal high SNR for all SN-CH and CH-FC is shown in Fig.\,\subref*{fig:ROC-SNR-Eq-hi}, in which the LLR and the LFR (employing equal power allocation, $P_m=1\, \forall m$) achieve the optimal performance provided by the OCR. By contrast, Fig.\,\subref*{fig:ROC-SNR-Eq-lo}  illustrates the case of having equal low SNR for all the clusters' channels. Here, the LLR rule performance is as good as that of the CR, whereas the LFR rule performs worse than the latter. This behaviour is explained by the direct dependence on the SNR in the weighing coefficients, in \eqref{eq:dm}. Fig.\,\subref*{fig:ROC-SNR-Cm-gt-Fm} shows the case of having better SN-CH channels compared to the CH-FC channels, while Fig.\,\subref*{fig:ROC-SNR-Fm-gt-Cm} shows the opposite case. The LFR performance virtually does not change whereas the LLR slightly degrades in the latter case. The LFR behaviour is explained by noting that $\sigmat^2_m$ is the weighted sum of $\sigma^2_{f,m}$ and $\sigma^2_{c,m}$. The LLR sensitivity might be attributed to its nonlinear form.

Next, Fig.\,\ref{fig:ROC-SNR-Target} shows the effect of having a good channel quality specifically in the cluster containing the target. Interestingly, the results here show that it is sufficient to have a good SN-CH channel quality in the cluster containing the target to achieve good detection performance as is evident in Fig.\,\subref*{fig:ROC-SNR-Target-hi-SNR_c} and Fig.\,\subref*{fig:ROC-SNR-Target-hi-SNR_f}. In a similar manner, a bad channel will significantly decrease the performance as can be seen in Fig.\,\subref*{fig:ROC-SNR-Target-lo-SNR_c} and Fig.\,\subref*{fig:ROC-SNR-Target-lo-SNR_f}. Fig.\,\ref{fig:Tail-Prob-Bnd} depicts the upper bounds for the $P_{FA}$ and the $P_D$ for the LFR presented in \eqref{eq:Thrm-Bnd}. 

Then the performance of the LFR-aML is compared with the LLR, LFR and the CR in Figure \ref{fig:ROC_PWR}. To make the comparison fair, all the rules use the received data average value in \eqref{eq:z_m_ave}. In Figure \ref{fig:ROC_aML}, the LFR-aML shows a satisfactory performance despite some loss of performance due to the inaccurate $\lambdaOm$ estimation. On the other hand, Figure \ref{fig:PWR_aML} shows the power allocation via \eqref{eq:Pm} over the CHs. Notice that with exact $\lambdaOm$ values, the power is concentrated in the cluster containing the target. Other clusters may receive no power allocation at all. However, using estimated values leads to spreading the power across the cluster, while giving more power to the target's cluster. 

Figure \ref{fig:PD_K1_FC_CH_2} shows $P_D$ using the LFR-PA plotted with respect to $D_1$, which is the detection performance constraint. The $P_D$ values are fitted with a third degree polynomial to emphasize the trend in a better way. In Figure \ref{fig:Psav_K1_FC_CH_2}, the power saving using the LFR-PA rule is plotted against $D_1$ for different WSN conditions. The power saving is defined as 
\begin{equation}
P_{\text{sav}} = \frac{1}{P_{tot}}\SumM \left(P_{tot} - P_m\right)\times 100\%
\end{equation}
where $P_{tot}$ is the power used by the LFR rule. The OCR, LFR, and CR are plotted for the sake of comparison. $P_D$ and $P_{\text{sav}}$ are plotted against $D_1$. It is clear that as $D_1$ increases, improved detection performance is achieved (in the range of 8\%) at the expense of less saved energy. However, the power saving is significantly greater in the case of a better FC-CH channel as shown. Indeed, at $D_1=5.5$ the power saving is 84\% for the case of $\SNR_{f,m}=5$dB and $\SNR_{c,m}=2$dB. Whereas it is 67\% for both cases $\SNR_{f,m}=\SNR_{c,m}=5$dB and $\SNR_{f,m}=\SNR_{c,m}=2$dB. Furthermore, the performance of the equal power version is attained with 64\% power saving. This follows from the direct proportionality of $P_m$ with $\sigma^2_{f,m}$ in \eqref{eq:Pm}.

\section{Conclusion}
\label{sec:Conclusion}
In this paper we have discussed fusion rules for DD in clustered-WSNs with communication channels experiencing AWGN. We have derived the optimal log-likelihood ratio fusion rule that turned out to be analytically intractable. A suboptimal linear fusion rule is subsequently derived, in which the cluster's data is linearly weighed. This rule, intuitively, gives more weight to clusters with better channels and better data quality. However, the LFR requires the knowledge of the mean value of detecting SNs. Thus we proposed the LFR-aML that employs an approximate constrained ML estimator to find the required parameters. 

In addition, Gaussian-tail upper bounds for the LFR's detection and false probabilities are derived using approximated moment generating functions. Moreover, a power allocation strategy is proposed that minimizes the total power used by the WSN. The resulting allocated power is proportional to the expected number of detecting SNs in the cluster. 

Extensive simulations show that in order to achieve near optimal detection performance, it is sufficient to have a good channel quality for the cluster(s) containing the detecting SNs and moderate quality in the rest of the clusters. However, when using the LFR-aML, there is a performance gap with the ideal LFR due to the inherent estimation errors. It has been shown that the proposed power allocation strategy can achieve 84\% power saving with only 5\% performance reduction compared to the equal power scheme. Furthermore, the same detection performance as the equal power allocation version can be attained with a  14\% power saving.

So in summary, for the first time a fusion rule has been derived for clustered WSN distributed detection with imperfect channels, and we have saved significant power usage with only a small reduction in detection performance. Several directions for future work can be pursued, including non-homogeneous PPP model and other types of channel imperfections such as path-loss, fading and channel failure. Furthermore, fusion rules for distributed detection can be investigated in which a decode-and-forward scheme (quantization) is employed  at the CH.
%
%
%
%
\appendices
\section{Proof of Lemma \eqref{lem:Bound}}
\label{app:App-Lemma}
Recall that the likelihood function of the $l$th received sample from the $m$th cluster under $\Hyp_1$ is actually the expectation of the conditional distribution $ p\left( \widetilde{z}_{l,m} ; \lambda_{1,m} \right) = \E_{\Lambda_m} \left[ p\left( \widetilde{z}_{l,m} \vert \Lambda_m \right) \right]$. Since $\Lambda_m$ is a Poisson RV, then the previous expectation becomes as follows:
\begin{equation}
 p\left( \widetilde{z}_{l,m} ; \lambda_{1,m} \right) = \sum_{k=0}^{\infty}  p\left( \widetilde{z}_{l,m} | k \right)  p\left( k ; \lambda_{1,m} \right)
\label{eq:app-lr}
\end{equation}
where $\Lambda_m$ is replaced by $k$ in order to simplify the notation. Note that \eqref{eq:app-lr} is a convex sum in terms of $p\left( \widetilde{z}_{l,m} | \Lambda_m \right)$ since $p\left( k ; \lambda_{1,m} \right)$ is a proper (Poisson) distribution\footnote{For any probability mass function all the probabilities must be less than one and also sum to unity.}. A lower bound of  $\log \left( p\left( \widetilde{z}_{l,m} ; \lambda_{1,m} \right) \right)$ can be attained via Jensen's inequality, however it is not useful in its current form. Instead, we proceed by treating the sum in \eqref{eq:app-lr} as a convex combination of $p\left( k ; \lambda_{1,m} \right)$ where $p\left( \widetilde{z}_{l,m} |k \right)$ are the weighing coefficients. But first, we need to normalize those coefficients as follows:


\begin{equation}
p\left( \widetilde{z}_{l,m} ; \lambda_{1,m} \right) = C_0 \sum_{k=0}^{\infty} \pi_k
  p\left( k ; \lambda_{1,m} \right)
\end{equation}
where $C_0 = \sum_{k=0}^{\infty}  p\left( \widetilde{z}_{l,m} | k \right)$ and $\pi_k = p\left( \widetilde{z}_{l,m} | k \right) /C_0$ is a discrete distribution. Now we are ready to apply Jensen's inequality on the log of \eqref{eq:app-lr} yielding
\begin{eqnarray}
\log \left( p\left( \widetilde{z}_{l,m} ; \lambda_{1,m} \right) \right) &\geq& \sum_{k=0}^{\infty} \pi_k \log p\left( k ; \lambda_{1,m} \right) + \log C_0 \nonumber \\
&=& \sum_{k=0}^{\infty} k \pi_k \log \lambda_{1,m} - \lambda_{1,m} \pi_k + C_1 \nonumber \\
&=& \widehat{\Lambda}_{l,m} \log \lambda_{1,m} - \lambda_{1,m} + C_1
\end{eqnarray}
where the second line above follows from the definition of the Poisson distribution and $C_1$ is a constant including the terms independent of $\lambda_{1,m}$. The last line results from the definition of the $\widehat{\Lambda}_{l,m}$ in \eqref{eq:zm_hat} and the fact that $\pi_k$'s sum up to unity.
%
%

\section{Proof of Lemma \eqref{lem:const-opt}}
\label{app:App-Lemma-const-opt}

First, problem \eqref{eq:app-ML} is put in the following canonical form:

\begin{eqnarray}
\label{eq:min-app-ML}
&&\min_{\lambdam} \, \sum_{m=1}^M \left(\lambdam  -  \widehat{\Lambda}_{m} \log\lambdam \right) \\
 &&\text{s.t.} \quad  \lambda_0 - \lambdam \leq 0 \quad \forall m. \nonumber
\end{eqnarray}
where the problem \eqref{eq:min-app-ML} is scaled by $1/L$, and $\widehat{\Lambda}_m$ is defined in \eqref{eq:Lambda-ave}. The corresponding Lagrangian is
\begin{equation}
\mathcal{L} = \sum_{m=1}^M  \lambdam - \widehat{\Lambda}_{m}\log\lambdam   + \eta_m \left( \lambda_0 - \lambdam \right)
\end{equation}
where $\eta_m \geq 0 \; \forall m$ are the slack variables. The corresponding KKT conditions \cite{Boyd2004} are
\begin{equation}
\frac{\partial \mathcal{L} }{\partial \lambdam} = 1 - \frac{\widehat{\Lambda}_{m}}{\lambda_{1,m}} - \eta_m = 0 \label{eq:derv_L}
\end{equation}
for all $m$ and the slack conditions are
\begin{equation}
\eta_m \left( \lambda_{1,m} - \lambda_0 \right) = 0 
\label{eq:comp-slack}
\end{equation}
with $\eta_m \geq 0$. The complementary slack conditions in \eqref{eq:comp-slack} dictate that $\lambda_{1,m} = \lambda_0$ when $\eta_m > 0$. Solving for the latter yields $\eta_m = 1 - \widehat{\Lambda}_m/\lambda_0$. Alternatively, when $\eta_m = 0$, that implies $\lambda_{1,m} = \widehat{\Lambda}_m$.

\section{Proof of Theorem \eqref{thrm:Noisy-Mod-Chrnf-Bound}}
\label{app:App-Noisy-Mod-Chrnf-Bound}

The MGF of $\LLFR$ in \eqref{eq:LFR} under either hypotheses is given by the conditional independence as 
\begin{eqnarray}
M_{\text{LFR}} &=& \prod_{m=1}^M M_{Z_m} \left(t d_m ; \Hyp_j \right) \nonumber \\
  		  &=& \prod_{m=1}^M M_{\Lambda_m} \left(t d_m \sqrt{\Pt_m} ; \Hyp_j  \right) M_{V_m} (t d_m).
\end{eqnarray}

From the MGFs of the Gaussian and Poison distributions we have
\begin{eqnarray}
M_{\text{LFR}}  		  
          &=& \exp \left( \sum_{m=1}^M \lambda_{j,m} \left( e^{t d_m \sqrt{\Pt_m}} - 1\right) + \frac{d^2_m \sigmat^2_m t^2}{2} \right). \nonumber \\
\end{eqnarray}

Using the second-order Taylor series for the exponential function the MGF becomes
{\small
\begin{eqnarray}
M_{\text{LFR}} &\approx& \exp \left( t \SumM \lambda_{j,m} d_m \sqrt{\Pt_m} + \frac{t^2}{2} d^2_m \left( \lambda_{j,m} \Pt_m + \sigmat^2_m \right) \right) \nonumber \\
           &=& \exp \left( \lambdaB_{j,d} t + \frac{\sigmat^2_{j,d}}{2} t^2 \right),
\end{eqnarray}
}
where $\lambdaB_{j,d}$ and $\sigmat^2_{j,d}$ are defined in \eqref{eq:lambda_bar} and \eqref{eq:sigma2_bar} respectively. The Chernoff bound for $\LLFR$ is
\begin{eqnarray}
\Prob \left( \LLFR > z \right) &<& \inf_{t>0} \exp\left(-zt\right) M_{\text{LFR}}(t) \nonumber \\
&=& \inf_{t>0} \exp \left( -zt + \lambdaB_{j,d} t + \frac{\sigmat^2_{j,d}}{2} t^2\right).
\end{eqnarray}

The infimum is found by taking the derivative of the exponential argument, equating to zero and solving for $t$. This gives the upper bound
\begin{equation}
\Prob \left( \LLFR > z; \Hyp_j \right) < \exp \left( - \frac{\left( z - \lambdaB_{j,d} \right)^2}{2\sigmat^2_{j,d}} \right).
\end{equation}

\section{Proof of Theorem \eqref{thrm:Opt-pwr-alloc}}
\label{app:Opt-pwr-alloc}

The Lagrangian of canonical optimization problem is
\begin{eqnarray}
\mathcal{L}\left( P_m,\, \mu_m \right) &=& \SumM P_m - \SumM \mu_m P_m \nonumber \\
    &+& \nu \left( D_1 - \SumM \frac{ P_m \lambda_{d,m}^2}{\sigma^2_{c,m} P_m + \sigma^2_{f,m}} \right)
\end{eqnarray}
where $\lambda_{d,m} = \left( \lambda_{1,m} - \lambda_0  \right)$ and $D_1 = D_0 /P_0$. Then the related KKT conditions are
\begin{eqnarray}
1 - \mu_m -\frac{\nu \lambda^2_{d,m} \sigma^2_{f,m} }{ \left( \sigma^2_{c,m} P_m + \sigma^2_{f,m} \right)^2 } &=& 0 \label{eq:Lag-derv} \\
\mu_m P_m &=& 0 \label{eq:slack-cond} \\
\nu \left( \SumM \frac{ P_m \lambda_{d,m}^2}{\sigma^2_{c,m} P_m + \sigma^2_{f,m}} - D_1 \right) &=& 0 \label{eq:MD-cond} \\
\mu_m \geq 0, \quad \nu & \geq & 0
\end{eqnarray}
for all $m$, where $\mu_m$'s and $\nu$ are the slack variables. To solve for $P_m$, we first recognize that if $\nu > 0$ then having $\mu_m > 0$ leads to $P_m = 0$ that follows from \eqref{eq:slack-cond}, which violates the condition \eqref{eq:MD-cond}. Hence, this forces $\mu_m = 0$ and $\nu > 0$. The solution of $P_m$ follows from the condition \eqref{eq:Lag-derv} as stated in \eqref{eq:Pm}. Finally, substituting the latter equation in \eqref{eq:MD-cond} gives $\nu$ solution stated in \eqref{eq:nu}.


\ifCLASSOPTIONcaptionsoff
  \newpage
\fi



%
\bibliographystyle{IEEEtran}
\bibliography{../../BibTeX_Data_Base/Main_Database}

\end{document}